CEBAF

The Continuous Electron Beam Accelerator Facility

Theory Group Preprint Series

Additional copies are available from the authors.

# Heavy Mesons In A Relativistic Model

J. Zeng, J. W. Van Orden and W. Roberts *

Department of Physics
Old Dominion University, Norfolk, VA 23529
and
Continuous Electron Beam Accelerator Facility
12000 Jefferson Avenue, Newport News, VA 23606

## Abstract

Motivated by the present interest in the heavy quark effective theory, we use the spectator equation to treat the mesonic bound states of heavy quarks. The kernel we use is based on scalar confining and vector Coulomb potentials. Wave functions are treated to leading order and energies to order $1/m_Q$ in the heavy-light systems, and order $1/m_Q^2$ in heavy-heavy systems. Our results are in reasonable agreement with experimental measurements. We estimate two of the parameters of the heavy quark effective theory, and propose further calculations that may be undertaken in the future.

*National Young Investigator

# Heavy Mesons In A Relativistic Model

J. Zeng, J. W. Van Orden and W. Roberts[†]
*Department of Physics, Old Dominion University, Norfolk, VA 23529*
*and*
*Continuous Electron Beam Accelerator Facility*
*12000 Jefferson Avenue, Newport News, VA 23606.*

Motivated by the present interest in the heavy quark effective theory, we use the spectator equation to treat the mesonic bound states of heavy quarks. The kernel we use is based on scalar confining and vector Coulomb potentials. Wave functions are treated to leading order and energies to order $1/m_Q$ in the heavy-light systems, and order $1/m_Q^2$ in heavy-heavy systems. Our results are in reasonable agreement with experimental measurements. We estimate two of the parameters of the heavy quark effective theory, and propose further calculations that may be undertaken in the future.

## I. INTRODUCTION

Recently, there has been great theoretical interest in hadrons containing $b$ and $c$ quarks. This has stemmed largely from the realization that, in the formal limit when the mass of one of the quarks in a hadron is taken to infinity, symmetries above and beyond those usually associated with quantum chromodynamics (QCD) arise. This realization has led to the development of the heavy quark effective theory (HQET) [1] [2] [3]. In the framework of this effective theory, corrections to the formal limit can be systematically included. One very important phenomenological consequence of this has been a number of attempts to extract $V_{cb}$ from experimental data, with little model dependence in the result.

Despite the power inherent in HQET, there is still much that this effective theory can not tell us about the properties of heavy hadrons. As an example, HQET allows us to infer the absolute normalization of some of the form factors necessary for describing the decays of hadrons with beauty to those with charm. We also know how to include, in a systematic way, corrections to these normalizations due to the finite masses of the $b$ and $c$ quarks, as well as those due to perturbative QCD effects. We can even deduce bounds on the slopes of these form factors at a particular kinematic point. However, we know nothing about the exact dependence of these form factors on kinematic invariants. As a second example, HQET leads us to the conclusion that the spectra of $B$ and $D$ mesons should be very much alike, modulo $1/m_b$ and $1/m_c$ effects. However, this effective theory tells us nothing about the details of the spectra, such as the exact ordering of states, or their masses. In essence, HQET provides a framework for systematically extracting symmetry relations and the corrections to the formal heavy-quark limit but can predict neither the spectra of the heavy mesons nor the approach to the heavy-quark limit. Until we know how to solve non-perturbative QCD, the details mentioned above, along with many others, are the realm of models: such models continue to play a crucial role in our understanding of QCD.

A model that is quite successful in predicting the mesonic spectra is the relativised constituent quark model of Godfrey and Isgur [4]. Indeed, it was this model and its applications to weak decays that originally suggested the existence of heavy-quark symmetries which in turn led to HQET. This model provides relativistic kinematic corrections to the standard nonrelativistic quark model using a linear confining potential and a color Coulomb interaction. Meson spectra calculated with this model are remarkably close to experimental masses in all flavor sectors. However, since one of the objectives of heavy quark theory is the calculation of weak decay amplitudes and form factors, it is necessary to use a relativistically covariant model.

A covariant extension to the Godfrey-Isgur model can be constructed using the spectator or Gross equation [5], which has been used with some success in models of the nucleon-nucleon interaction [6], as well as in quark models of mesons composed of equal mass quarks and antiquarks [7]. This equation can be related to the Bethe-Salpeter equation by placing one of the intermediate-state particles on the positive-energy mass-shell. This has the advantages that the prescribed constraint on the relative energy is manifestly covariant and that in the limit that the mass of one constituent goes to infinity (the static limit), the wave equation reduces to the Dirac equation for the light particle [8]. This is a property of the full Bethe-Salpeter equation that is lost when the infinite sum of contributions to the kernel is truncated. Clearly, the properties of the spectator equation make it ideal for studying the properties of heavy mesons at finite mass.

In this article we use the spectator equation to construct a constituent quark model of heavy mesons. In particular, we will use the spectator equation as a basis for construction and expansion of the heavy meson spectra and wave functions in $1/m_Q$, where $m_Q$ is the heavy quark mass. This allows us to study the heavy meson spectra in the approach to the heavy quark symmetry limit. By choosing a reasonable set of model parameters we are able to obtain a respectable fit to the observed heavy meson masses and to predict the approximate masses of heavy mesons which have not yet been observed.

This article is organized as follows. In the next section, we describe the model that we use for heavy mesons, including the derivation of a wave equation from the spectator equation. In Section III, three methods of obtaining solutions of the wave equation are described, while in Section IV we display our results. In Section V, we present some conclusions.

## II. THE MODEL

### A. $Q\bar{q}$ and $q\bar{Q}$ mesons

The spectator equation is most easily understood in relation to the Bethe-Salpeter equation. The Bethe-Salpeter vertex function for two bound fermions is represented by Fig. 1 and can be written as

$$\Gamma(p,P) = i\int \frac{d^4k}{(2\pi)^4} V(p,k;P) S_F^{(1)}(k_1,m_1) S_F^{(2)}(k_2,m_2)\Gamma(k,P), \qquad (1)$$

where $p = \frac{1}{2}(p_1 - p_2)$, $k = \frac{1}{2}(k_1 - k_2)$, $V$ is the Bethe-Salpeter kernel and $S_F^{(i)}(k_i,m_i)$ is the free Dirac propagator for particle $i$. The Dirac indices are suppressed for simplicity.

The spectator vertex function can be obtained from the Bethe-Salpeter vertex function by placing one of the fermions on its positive-energy mass-shell. For our model the heavy quark (particle 2) is placed on shell while the light quark (particle 1) remains off shell. This is achieved by a replacement of the propagator

$$S_F^{(2)}(k_2,m_2) \to -2\pi i \frac{m_2}{E(\mathbf{k_2},m_2)}\delta\left(k^0 - \frac{P^0}{2} + E(\mathbf{k_2},m_2)\right)\Lambda^{+(2)}(k_2,m_2), \qquad (2)$$

where

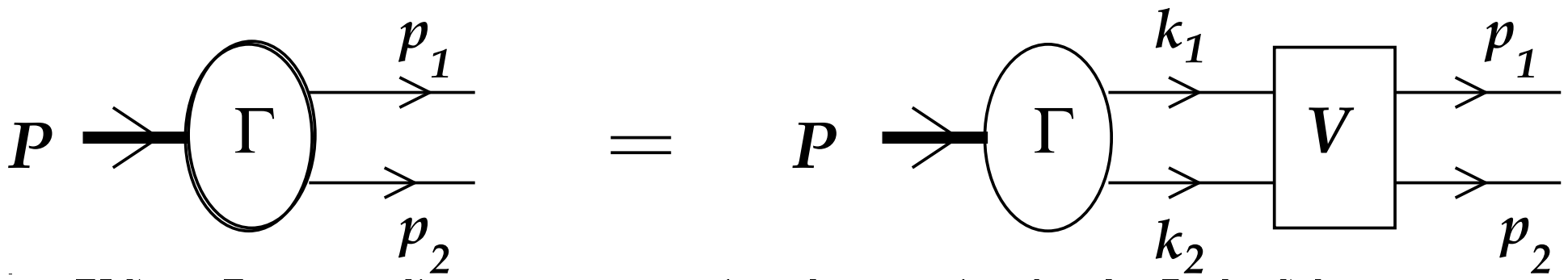

FIG. 1. Feynman diagrams representing the equation for the Bethe-Salpeter vertex function.

$$\Lambda^{+(2)}(k_2,m_2) = \sum_{s_2'} u^{(2)}(\mathbf{k_2},s_2',m_2)\bar{u}^{(2)}(\mathbf{k_2},s_2',m_2), \qquad (3)$$

and replacing $p$, $k$ and $k_1$ by the corresponding quantities $\check{p}$, $\check{k}$ and $\check{k}_1$ with particle 2 on mass shell. The on-shell energy is given by $E(\mathbf{p},m) = \sqrt{\mathbf{p}^2 + m^2}$. The spectator vertex function is then

$$\Gamma(\check{p},P) = \int \frac{d^3k}{(2\pi)^3}\frac{m_2}{E(\mathbf{k_2},m_2)} V(\check{p},\check{k};P) S_F^{(1)}(\check{k}_1,m_1)\Lambda^{+(2)}(k_2,m_2)\Gamma(\check{k},P). \quad (4)$$

Defining the spectator wave function as

$$\psi_{s_2}(\check{p},P) = S_F^{(1)}(\check{p}_1,m_1)\bar{u}^{(2)}(\mathbf{p_2},s_2,m_2)\Gamma(\check{p},P), \qquad (5)$$

the wave function satisfies the wave equation

$$S_F^{(1)^{-1}}(\check{p}_1,m_1)\psi_{s_2}(\check{p},P) = \int \frac{d^3k}{(2\pi)^3}\frac{m_2}{E(\mathbf{k_2},m_2)}\sum_{s_2'} V_{s_2,s_2'}(\check{p},\check{k};P)\psi_{s_2'}(\check{k},P), \qquad (6)$$

where

$$V_{s_2,s_2'}(\check{p},\check{k};P) = \bar{u}^{(2)}(\mathbf{p_2},s_2,m_2)V(\check{p},\check{k};P)u^{(2)}(\mathbf{k_2},s_2',m_2). \qquad (7)$$

This wave equation is covariant and can be easily boosted from frame to frame. It is generally easier to solve the wave equation in the bound-state rest frame where the angular expansions of the wave function and potential are defined. In the rest frame $P = (W,\mathbf{0})$, $\mathbf{p_1} = -\mathbf{p_2} = \mathbf{p}$, $\mathbf{k_1} = -\mathbf{k_2} = \mathbf{k}$, $p_1^0 = W - E(\mathbf{p},m_2)$, $p_2^0 = E(\mathbf{p},m_2)$, $k_1^0 = W - E(\mathbf{k},m_2)$, and $k_2^0 = E(\mathbf{k},m_2)$ where $W$ is the bound-state mass. The wave equation can be written as

$$\left[\gamma^{(1)^0}(W - E(\mathbf{p},m_2)) - \boldsymbol{\gamma}^{(1)}\cdot\mathbf{p} - m_1\right]\psi_{s_2}(\mathbf{p},W) = \int \frac{d^3k}{(2\pi)^3}\frac{m_2}{E(\mathbf{k},m_2)}\sum_{s_2'} V_{s_2,s_2'}(\mathbf{p},\mathbf{k};W)\psi_{s_2'}(\mathbf{k},W), \qquad (8)$$

where

$$V_{s_2,s_2'}(\mathbf{p},\mathbf{k};W)=\bar{u}^{(2)}(-\mathbf{p},s_2,m_2)V(\mathbf{p},\mathbf{k};W)u^{(2)}(-\mathbf{k},s_2',m_2). \qquad (9)$$

Since we wish to examine the approach to the limit $m_2 \to \infty$, it is useful to rewrite this equation in a noncovariant form by defining

$$\Psi_{s_2}(\mathbf{p}) \equiv \sqrt{\frac{m_2}{E(\mathbf{p},m_2)}}\psi_{s_2}(\mathbf{p},W) \qquad (10)$$

and

$$U_{s_2,s_2'}(\mathbf{p},\mathbf{k};W) \equiv \sqrt{\frac{m_2}{E(\mathbf{p},m_2)}}V_{s_2,s_2'}(\mathbf{p},\mathbf{k};W)\sqrt{\frac{m_2}{E(\mathbf{k},m_2)}} \qquad (11)$$

to give

$$\left[\gamma^{(1)^0}(W-E(\mathbf{p},m_2))-\boldsymbol{\gamma}^{(1)}\cdot\mathbf{p}-m_1\right]\Psi_{s_2}(\mathbf{p})=$$
$$\int\frac{d^3k}{(2\pi)^3}\sum_{s_2'}U_{s_2,s_2'}(\mathbf{p},\mathbf{k};W)\Psi_{s_2'}(\mathbf{k}). \qquad (12)$$

It is necessary to assume some form for the kernel $V$ in order to expand about the infinite mass limit. Here we assume that the kernel is of the simplest form which can be reduced to that used in ref. [4]. We choose the kernel to be

$$V(p,k;P)=V_s(Q^2)+\gamma^{(1)}\cdot\gamma^{(2)}V_v(Q^2), \qquad (13)$$

where

$$Q^2=(\mathbf{k}-\mathbf{p})^2-\left[E(\mathbf{k},m_2)-E(\mathbf{p},m_2)\right]^2. \qquad (14)$$

$V_v(Q^2)$ is a vector potential which is a color Coulomb interaction and the confining force is the result of the scalar potential $V_s(Q^2)$. This choice of interaction assumes that the Lorentz gauge is used in the color Coulomb interaction.

Using the explicit form of the Dirac spinors in (12) and the Dirac $\gamma$-matrices to reduce particle 2 to the Pauli spin space, and defining a wave function which is an operator in the Dirac space of particle 1 and the Pauli space of particle 2, $\Psi=\sum_{s_2'}\chi_{s_2'}\Psi_{s_2'}$, (12) becomes

$$\left(\gamma^{(1)^0}(W-E(\mathbf{p},m_2))-\boldsymbol{\gamma}^{(1)}\cdot\mathbf{p}-m_1\right)\Psi(\mathbf{p})=$$
$$\int\frac{d^3k}{(2\pi)^3}\left(\frac{(E(\mathbf{p},m_2)+m_2)(E(\mathbf{k},m_2)+m_2)}{4E(\mathbf{p},m_2)E(\mathbf{k},m_2)}\right)^{\frac{1}{2}}$$
$$\times\left\{\left(1-\frac{\boldsymbol{\sigma}^{(2)}\cdot\mathbf{p}\boldsymbol{\sigma}^{(2)}\cdot\mathbf{k}}{(E(\mathbf{p},m_2)+m_2)(E(\mathbf{k},m_2)+m_2)}\right)V_s(Q^2)\right.$$
$$+\gamma^{(1)^0}\left(1+\frac{\boldsymbol{\sigma}^{(2)}\cdot\mathbf{p}\boldsymbol{\sigma}^{(2)}\cdot\mathbf{k}}{(E(\mathbf{p},m_2)+m_2)(E(\mathbf{k},m_2)+m_2)}\right)V_v(Q^2)$$
$$\left.+\boldsymbol{\gamma}^{(1)}\cdot\left(\frac{\boldsymbol{\sigma}^{(2)}\cdot\mathbf{p}\boldsymbol{\sigma}^{(2)}}{(E(\mathbf{p},m_2)+m_2)}+\frac{\boldsymbol{\sigma}^{(2)}\boldsymbol{\sigma}^{(2)}\cdot\mathbf{k}}{(E(\mathbf{k},m_2)+m_2)}\right)V_v(Q^2)\right\}\Psi(\mathbf{k}), \qquad (15)$$

Expanding eq. (15) to order $1/m_2$, we find

$$\left(\gamma^{(1)^0}(W-m_2-\frac{\mathbf{p}^2}{2m_2})-\boldsymbol{\gamma}^{(1)}\cdot\mathbf{p}-m_1\right)\Psi(\mathbf{p})=$$
$$\int\frac{d^3k}{(2\pi)^3}\left(V_s(\mathbf{q}^2)+\gamma^{(1)^0}V_v(\mathbf{q}^2)\right.$$
$$\left.+\frac{1}{2m_2}\boldsymbol{\gamma}^{(1)}\cdot(\boldsymbol{\sigma}^{(2)}\boldsymbol{\sigma}^{(2)}\cdot\mathbf{k}+\boldsymbol{\sigma}^{(2)}\cdot\mathbf{p}\boldsymbol{\sigma}^{(2)})V_v(\mathbf{q}^2)\right)\Psi(\mathbf{k}). \qquad (16)$$

where $\mathbf{q}=\mathbf{k}-\mathbf{p}$.

Eq. (16) can be Fourier transformed to coordinate space, multiplied from the left by $\gamma^{(1)^0}$ and then rearranged to give the wave equation

$$H\Psi(\mathbf{r})=W\Psi(\mathbf{r}), \qquad (17)$$

where the hermitian hamiltonian is $H=H_0+H_1$ with

$$H_0=\boldsymbol{\alpha}^{(1)}\cdot\frac{1}{i}\boldsymbol{\nabla}+\beta^{(1)}m_1+\beta^{(1)}V_s(r)+V_v(r)+m_2, \qquad (18a)$$

$$H_1=\frac{1}{2m_2}\left\{-\nabla^2-i\left\{V_v(r),\boldsymbol{\alpha}^{(1)}\cdot\boldsymbol{\nabla}\right\}+\boldsymbol{\alpha}^{(1)}\cdot\boldsymbol{\sigma}^{(2)}\times\hat{\mathbf{r}}V_v'(r)\right\}, \qquad (18b)$$

where $\hat{r}$ is the unit vector in the radial direction.

Eq. (18a) is the Dirac equation for particle 1 with scalar and vector potentials plus the mass of the heavy quark, particle 2. The solutions of the Dirac equation with such a potential have been extensively studied. The operators

$$\left\{\mathbf{j}^{(1)^2},j_z^{(1)},K,S_z^{(2)}\right\} \qquad (19)$$

TABLE I. Values of $\ell$ and $\bar{\ell}$ for various values of $\kappa$

| | $\ell$ | $\bar{\ell}$ |
|---|---|---|
| $\kappa_1 < 0$ | $j_1 - \frac{1}{2}$ | $j_1 + \frac{1}{2}$ |
| $\kappa_1 > 0$ | $j_1 + \frac{1}{2}$ | $j_1 - \frac{1}{2}$ |

are a set of mutually commuting operators which commute with $H_0$, where $\mathbf{j}^{(1)} = \mathbf{L} + \mathbf{S}^{(1)}$, $\mathbf{S}^{(1)} = \frac{1}{2}\mathbf{\Sigma}^{(1)} = \frac{1}{2}\gamma_5^{(1)}\boldsymbol{\alpha}^{(1)}$, $K^{(1)} = \beta^{(1)}(\mathbf{\Sigma}^{(1)} \cdot \mathbf{j}^{(1)} - \frac{1}{2})$ and $\mathbf{S}^{(2)} = \frac{1}{2}\boldsymbol{\sigma}^{(2)}$. The eigenstates of $H_0$ can then be labelled by the corresponding set of quantum numbers $\{n, j_1, m_{j_1}, \kappa_1, s_2\}$. The wave equation associated with $H_0$ can then be written as

$$H_0 \Psi^{(0)}_{n\kappa_1 j_1 m_{j_1} s_2}(\mathbf{r}) = W^{(0)}_{n\kappa_1 j_1} \Psi^{(0)}_{n\kappa_1 j_1 m_{j_1} s_2}(\mathbf{r}), \tag{20}$$

where

$$\Psi^{(0)}_{n\kappa_1 j_1 m_{j_1} s_2}(\mathbf{r}) = \begin{pmatrix} \frac{G_{n\ell j_1}(r)}{r} \mathcal{Y}^{m_{j_1}}_{\ell \frac{1}{2} j_1}(\Omega) \\ \frac{iF_{n\ell j_1}(r)}{r} \mathcal{Y}^{m_{j_1}}_{\bar{\ell} \frac{1}{2} j_1}(\Omega) \end{pmatrix} \chi_{s_2}, \tag{21}$$

with

$$\mathcal{Y}^{m_{j_1}}_{\ell \frac{1}{2} j_1}(\Omega) = \sum_{m_\ell, s_1} \left\langle \ell m_\ell, \frac{1}{2} s_1 \middle| j_1 m_{j_1} \right\rangle Y_{\ell m_\ell}(\Omega) \chi_{s_1}, \tag{22}$$

and $\chi_{s_1}$ and $\chi_{s_2}$ are the Pauli spinors for particles 1 and 2, respectively. The eigenvalue $\kappa_1 = \pm(j_1 + \frac{1}{2})$ can be any nonzero integer. The values of $\ell$ and $\bar{\ell}$ associated with various values of $\kappa_1$ are displayed in Table I.

Note that the zeroth order invariant mass $W^{(0)}_{n\kappa_1 j_1}$ is determined by $n$, $\kappa_1$ and $j_1$, or equivalently by $n$, $j_1$, and $\ell$. The parity of the $Q\bar{q}$ bound state is given by $P = (-1)^{\ell+1}$.

The first term on the right hand side of (18b) is the kinetic energy of particle 2. Both the first and second terms on the right hand side of (18b) commute with the set of operators given in (19). However, the third term does not commute with any of these operators, but instead commutes with

$$\{\mathbf{J}^2, J_z, \mathcal{P}\} \tag{23}$$

where $\mathbf{J} = \mathbf{j}^{(1)} + \mathbf{S}^{(2)}$ and $\mathcal{P}$ is the parity operator. The eigenstates of the total hamiltonian $H = H_0 + H_1$ can then be labelled by the set of quantum numbers $\{n, J, M_J, P\}$.

The eigenstates and eigenenergies of the hamitonian $H$ can be calculated directly. However, the objective of the calculations presented here is to produce wave functions which can be used in the calculation of form factors and decay constants as an expansion in powers of the inverse of the heavy quark mass $m_2$. In order to maintain consistency in this expansion, the masses and wave functions should be calculated perturbatively. The first order correction to the quark bound state mass is given by

$$W^{(1)}_{nJP} = \int d^3r \Psi^{(0)\dagger}_{n\kappa_1 j_1 J M_J}(\mathbf{r}) H_1 \Psi^{(0)}_{n\kappa_1 j_1 J M_J}(\mathbf{r}), \tag{24}$$

where

$$\Psi^{(0)}_{n\kappa_1 j_1 J M_J}(\mathbf{r}) = \sum_{m_{j_1}, s_2} \left\langle j_1 m_{j_1}, \frac{1}{2} s_2 \middle| J M_J \right\rangle \Psi^{(0)}_{n\kappa_1 j_1 m_{j_1} s_2}(\mathbf{r}). \tag{25}$$

The bound state mass to first order is

$$W_{nJP} = W^{(0)}_{n\kappa_1 j_1} + W^{(1)}_{nJP}. \tag{26}$$

The scalar and vector potentials in the calculations presented here have the form

$$V_s(r) = br + c, \tag{27}$$

$$V_v(r) = -\frac{4}{3} \sum_{i=1}^{3} \frac{\alpha_i}{r} \mathrm{erf}(\gamma_i r). \tag{28}$$

The vector potential is, as in ref. [4], based on a parametrization of the running QCD coupling constant.

### B. $Q\bar{Q}$ mesons

The situation for mesons made of a heavy quark and the corresponding antiquark is somewhat more complicated. The problem is that the prescription of placing particle 2 on mass shell in the Bethe-Salpeter vertex equation (1) to obtain the spectator vertex equation (4) is clearly asymmetrical. This results in a spectator vertex function which is no longer an eigenfunction of the charge conjugation operator. The solution of this problem is to construct a set of coupled equations for the vertex functions which have either particle 1 or particle 2 on mass shell [7]. These equations have been solved in ref. [7] for $q\bar{q}$-systems containing only light quarks.

However, since we are interested in expanding about the infinite mass limit, this additional complication is not necessary and a hamiltonian with leading $1/m_Q$ corrections can be constructed from (4). The starting point is the spinor decomposition of the Dirac propagator of particle 1 in the meson rest frame

$$S_F^{(1)}(\check{k}_1, m_Q) = \frac{m_Q}{E(\mathbf{k}, m_Q)} \sum_{s_1'} \left[ \frac{u^{(1)}(\mathbf{k}, s_1', m_Q)\bar{u}^{(1)}(\mathbf{k}, s_1', m_Q)}{W - 2E(\mathbf{k}, m_Q) + i\eta} + \frac{v^{(1)}(-\mathbf{k}, s_1', m_Q)\bar{v}^{(1)}(-\mathbf{k}, s_1', m_Q)}{W - i\eta} \right]. \quad (29)$$

Using eqs. (29) and (3) in eq. (4), we can write [9]

$$\Gamma(\check{p}, P) = \sum_{s_1' s_2'} \int \frac{d^3k}{(2\pi)^3} \frac{m_Q}{E(\mathbf{k}, m_Q)} V(\check{p}, \check{k}; P) \left( u^{(1)}(\mathbf{k}, s_1', m_Q) u^{(2)}(-\mathbf{k}, s_2', m_Q) \Psi^{(+)}_{s_1', s_2'}(\mathbf{k}) + v^{(1)}(-\mathbf{k}, s_1', m_Q) u^{(2)}(-\mathbf{k}, s_2', m_Q) \Psi^{(-)}_{s_1', s_2'}(\mathbf{k}) \right), \quad (30)$$

where

$$\Psi^{(+)}_{s_1', s_2'}(\mathbf{k}) = \frac{m_Q}{E(\mathbf{k}, m_Q)} \frac{\bar{u}^{(1)}(\mathbf{k}, s_1', m_Q)\bar{u}^{(2)}(-\mathbf{k}, s_2', m_Q)\Gamma(\check{k}, P)}{W - 2E(\mathbf{k}, m_Q)}, \quad (31)$$

and

$$\Psi^{(-)}_{s_1', s_2'}(\mathbf{k}) = \frac{m_Q}{E(\mathbf{k}, m_Q)} \frac{\bar{v}^{(1)}(-\mathbf{k}, s_1', m_Q)\bar{u}^{(2)}(-\mathbf{k}, s_2', m_Q)\Gamma(\check{k}, P)}{W}. \quad (32)$$

Multiplying the terms of (30) to the left respectively by

$$\frac{m_Q}{E(\mathbf{p}, m_Q)} \bar{u}^{(1)}(\mathbf{p}, s_1, m_Q)\bar{u}^{(2)}(-\mathbf{p}, s_2, m_Q) \quad (33)$$

and

$$\frac{m_Q}{E(\mathbf{p}, m_Q)} \bar{v}^{(1)}(-\mathbf{p}, s_1, m_Q)\bar{u}^{(2)}(-\mathbf{p}, s_2, m_Q), \quad (34)$$

means that eq. (12) can be rewritten as the pair of coupled integral equations

$$(W - 2E(\mathbf{p}, m_Q)) \Psi^{(+)}_{s_1, s_2}(\mathbf{p}) = \sum_{s_1', s_2'} \int \frac{d^3k}{(2\pi)^3} \left[ U^{++}_{s_1, s_2; s_1', s_2'}(\mathbf{p}, \mathbf{k}; W) \Psi^{(+)}_{s_1', s_2'}(\mathbf{k}) + U^{+-}_{s_1, s_2; s_1', s_2'}(\mathbf{p}, \mathbf{k}; W) \Psi^{(-)}_{s_1', s_2'}(\mathbf{k}) \right], \quad (35)$$

and

$$W \Psi^{(-)}_{s_1, s_2}(\mathbf{p}) = \sum_{s_1', s_2'} \int \frac{d^3k}{(2\pi)^3} \left[ U^{-+}_{s_1, s_2; s_1', s_2'}(\mathbf{p}, \mathbf{k}; W) \Psi^{(+)}_{s_1', s_2'}(\mathbf{k}) + U^{--}_{s_1, s_2; s_1', s_2'}(\mathbf{p}, \mathbf{k}; W) \Psi^{(-)}_{s_1', s_2'}(\mathbf{k}) \right], \quad (36)$$

where

$$U^{++}_{s_1, s_2; s_1', s_2'}(\mathbf{p}, \mathbf{k}; W) = \frac{m_Q^2}{E(\mathbf{p}, m_Q)E(\mathbf{k}, m_Q)} \times \bar{u}^{(1)}(\mathbf{p}, s_1)\bar{u}^{(2)}(-\mathbf{p}, s_2) V(\mathbf{p}, \mathbf{k}; W) u^{(1)}(\mathbf{k}, s_1') u^{(2)}(-\mathbf{k}, s_2'), \quad (37)$$

$$U^{+-}_{s_1, s_2; s_1', s_2'}(\mathbf{p}, \mathbf{k}; W) = \frac{m_Q^2}{E(\mathbf{p}, m_Q)E(\mathbf{k}, m_Q)} \times \bar{u}^{(1)}(\mathbf{p}, s_1, m_Q)\bar{u}^{(2)}(-\mathbf{p}, s_2) V(\mathbf{p}, \mathbf{k}; W) v^{(1)}(-\mathbf{k}, s_1') u^{(2)}(-\mathbf{k}, s_2'), \quad (38)$$

$$U^{-+}_{s_1, s_2; s_1', s_2'}(\mathbf{p}, \mathbf{k}; W) = \frac{m_Q^2}{E(\mathbf{p}, m_Q)E(\mathbf{k}, m_Q)} \times \bar{v}^{(1)}(-\mathbf{p}, s_1)\bar{u}^{(2)}(-\mathbf{p}, s_2) V(\mathbf{p}, \mathbf{k}; W) u^{(1)}(\mathbf{k}, s_1') u^{(2)}(-\mathbf{k}, s_2'), \quad (39)$$

$$U^{--}_{s_1, s_2; s_1', s_2'}(\mathbf{p}, \mathbf{k}; W) = \frac{m_Q^2}{E(\mathbf{p}, m_Q)E(\mathbf{k}, m_Q)} \times \bar{v}^{(1)}(-\mathbf{p}, s_1, m_Q)\bar{u}^{(2)}(-\mathbf{p}, s_2) V(\mathbf{p}, \mathbf{k}; W) v^{(1)}(-\mathbf{k}, s_1') u^{(2)}(-\mathbf{k}, s_2'). \quad (40)$$

These coupled equations can then be reduced to the Pauli spin space and expanded in powers of $1/m_Q$. In this case, only $U^{++}\Psi^{(+)}$ contributes to order $1/m_Q^2$. Defining a wave function which is an operator in the spin spaces of both particles as

$$\Psi = \sum_{s_1', s_2'} \chi_{s_1'} \chi_{s_2'} \Psi^{(+)}_{s_1', s_2'}, \quad (41)$$

eq. (35) becomes

$$\left( W - 2m_Q - \frac{\mathbf{p}^2}{m_Q} \right) \Psi(\mathbf{p}) = \int \frac{d^3k}{(2\pi)^3} U(\mathbf{p}, \mathbf{k}) \Psi(\mathbf{k}), \quad (42)$$

where

$$U(\mathbf{p},\mathbf{k}) = V_s(\mathbf{q}^2) + V_v(\mathbf{q}^2) - \frac{1}{4m_Q^2}\Big[\left(V_s'(\mathbf{q}^2) + V_v'(\mathbf{q}^2)\right)\left(\mathbf{k}^2 - \mathbf{p}^2\right)^2$$
$$+V_s(\mathbf{q}^2)\left(\mathbf{p}^2 + \mathbf{k}^2 + \boldsymbol{\sigma}^{(1)}\cdot\mathbf{p}\boldsymbol{\sigma}^{(1)}\cdot\mathbf{k} + \boldsymbol{\sigma}^{(2)}\cdot\mathbf{p}\boldsymbol{\sigma}^{(2)}\cdot\mathbf{k}\right)$$
$$+V_v(\mathbf{q}^2)\left(\mathbf{p}^2 + \mathbf{k}^2 - \boldsymbol{\sigma}^{(1)}\cdot\mathbf{p}\boldsymbol{\sigma}^{(1)}\cdot\mathbf{k} - \boldsymbol{\sigma}^{(2)}\cdot\mathbf{p}\boldsymbol{\sigma}^{(2)}\cdot\mathbf{k}\right)$$
$$- V_v(\mathbf{q}^2)\left(\boldsymbol{\sigma}^{(1)}\boldsymbol{\sigma}^{(1)}\cdot\mathbf{k} + \boldsymbol{\sigma}^{(1)}\cdot\mathbf{p}\boldsymbol{\sigma}^{(1)}\right)\cdot\left(\boldsymbol{\sigma}^{(2)}\boldsymbol{\sigma}^{(2)}\cdot\mathbf{k} + \boldsymbol{\sigma}^{(2)}\cdot\mathbf{p}\boldsymbol{\sigma}^{(2)}\right)\Big]. \quad (43)$$

Eq. (42) can then be Fourier transformed to coordinate space to extract the hamiltonian

$$H = H_0 + H_1, \quad (44)$$

with

$$H_1 = H_{\rm c} + H_{\rm hyp} + H_{\rm so} + H_{\rm SR} + H_{\rm VR}, \quad (45)$$

where

$$H_0 = -\frac{\nabla^2}{m_Q} + V_s(r) + V_v(r) + 2m_Q, \quad (46a)$$

$$H_{\rm c} = \frac{1}{m_Q^2}\left\{\frac{1}{4}\left[\nabla^2 V_s(r)\right] - \left[V_v(r) - V_s(r)\right]\nabla^2 + \left[V_s'(r) - V_v'(r)\right]\frac{\partial}{\partial r}\right\}, \quad (46b)$$

$$H_{\rm hyp} = \frac{1}{m_Q^2}\left\{\frac{1}{2}\left[\frac{1}{r}V_v'(r) - V_v''(r)\right]\left(\mathbf{S}\cdot\hat{\mathbf{r}}\mathbf{S}\cdot\hat{\mathbf{r}} - \frac{1}{3}\mathbf{S}^2\right) + \left[\nabla^2 V_v(r)\right]\left(\frac{1}{3}\mathbf{S}^2 - \frac{1}{2}\right)\right\}, \quad (46c)$$

$$H_{\rm so} = \frac{1}{2m_Q^2 r}\left[3V_v'(r) - V_s'(r)\right]\mathbf{S}\cdot\mathbf{L}, \quad (46d)$$

$$H_{\rm S(V)R} = -\frac{1}{4m_Q^2}\left[\boldsymbol{\nabla}^2, \left[\boldsymbol{\nabla}^2, F_{\rm S(V)R}(\mathbf{x})\right]\right], \quad (46e)$$

and $\mathbf{S} = \mathbf{S}^{(1)} + \mathbf{S}^{(2)}$. Here $F_{\rm S(V)R}(\mathbf{x})$ is the Fourier transformation of $dV_{s(v)}(\mathbf{q}^2)/d\mathbf{q}^2$. For our choices of $V_s(r)$ and $V_v(r)$, we find

$$H_{\rm SR} = \frac{b}{m_Q^2}\left(\frac{\mathbf{L}^2}{2r} - 3\frac{\partial}{\partial r} - r\frac{\partial^2}{\partial r^2} - \frac{1}{r}\right) - \frac{c}{m_Q^2}\left(\frac{1}{r}\frac{\partial}{\partial r} + \frac{1}{2}\frac{\partial^2}{\partial r^2} - \frac{\mathbf{L}^2}{2r^2}\right), \quad (46f)$$

$$H_{\rm VR} = \frac{V_v(r)}{2m_Q^2 r^2}\mathbf{L}^2 - \frac{1}{3m_Q^2\sqrt{\pi}}\sum_i \alpha_i\gamma_i\, e^{-\gamma_i^2 r^2}\left(10\gamma_i^2 - 4\gamma_i^4 r^2 + 8\gamma_i^2 r\frac{\partial}{\partial r} - \frac{8}{r}\frac{\partial}{\partial r} - 4\frac{\partial^2}{\partial r^2}\right), \quad (46g)$$

Eq. (46a) is the nonrelativistic hamiltonian for equal mass quarks in scalar and vector potentials. $H_{\rm c}$ contains central and orbital contributions. $H_{\rm hyp}$ is the hyperfine interaction consisting of a tensor-force term and a spin-spin interaction. $H_{\rm so}$ is the spin-orbit interaction. $H_{\rm SR}$ and $H_{\rm VR}$ are scalar and vector retardation terms associated with the third term on the right-hand side of (43). Note that our spin-dependent interactions $H_{\rm hyp}$ and $H_{\rm so}$ have the same forms as those in many other quark models (see for example: [4,10,11]), but the spin-independent interactions do not.

The spin-independent correction includes $H_{\rm c}$, $H_{\rm SR}$ and $H_{\rm VR}$. In these contributions, $H_{\rm SR}$, $H_{\rm VR}$ and the term $[V_s'(r) - V_v'(r)]\frac{\partial}{\partial r}$ in $H_c$ are gauge dependent. $H_{\rm SR}$ and $H_{\rm VR}$ are from the second term in the expansion of $V(Q^2) = V(\mathbf{q}^2) - \frac{1}{4m_Q^2}V'(\mathbf{q}^2)\left(\mathbf{k}^2 - \mathbf{p}^2\right)^2 + \mathcal{O}(1/m_Q^3)$. Had we chosen the Coulomb gauge, these terms would not exist. Most other quark models do not include retarded interactions. (Ref. [12] gives another expression for the retardation effect.) We will show that with the scalar and vector potentials in (27) and (28), retardation contributions are comparable with the spin-dependent interactions.

The operators $\{H_0, \mathbf{L}^2, \mathbf{S}^2, \mathbf{J}^2, \mathbf{J}_z\}$ where $\mathbf{J} = \mathbf{L} + \mathbf{S}$, are a set of mutually commuting hermitian operators. The eigenstates of $H_0$ can then be labelled by the corresponding set of quantum numbers $\{n, L, S, J, M_J\}$. The wave equation associated with $H_0$ can then be written as

$$H_0\Psi^{(0)}_{nLSJM_J}(\mathbf{r}) = W^{(0)}_{nL}\Psi^{(0)}_{nLSJM_J}(\mathbf{r}), \quad (47)$$

where

$$\Psi^{(0)}_{nLSJM_J}(\mathbf{r}) = \frac{u_{nL}(r)}{r}\mathcal{Y}^{M_J}_{LSJ}(\Omega), \quad (48)$$

and

$$\mathcal{Y}_{LSJ}^{M_J}(\Omega) = \sum_{M_L, M_S} \langle LM_L SM_S \, | JM_J \rangle \, Y_{LM_L}(\Omega) \, |SM_S\rangle \tag{49}$$

is the spin spherical harmonic.

The hyperfine interaction (46c) mixes states with $\Delta L = \pm 2$ for $S = 1$. As a result, $L$ is no longer a good quantum number for solutions of the complete hamiltonian. However, these states have the same parity and charge quantum numbers since $P = (-1)^{L+1}$ and $C = (-1)^{L+S}$ for $\Psi^{(0)}$. The first-order correction to the mass can then be written as

$$\begin{aligned} W_{nJPC}^{(1)} &= \int d^3 r \Psi_{nLSJM_J}^{(0)\dagger}(\mathbf{r}) H_1 \Psi_{nLSJM_J}^{(0)}(\mathbf{r}) \\ &= E_{\rm c} + E_{\rm hyp} + E_{\rm so} + E_{\rm SR} + E_{\rm VR}. \end{aligned} \tag{50}$$

where $P = (-1)^{L+1}$ and $C = (-1)^{L+S}$. The bound state mass to first order is

$$W_{nJPC} = W_{nL}^{(0)} + W_{nJPC}^{(1)} \tag{51}$$

One may also include an annihilation term in the hamiltonian. However, this term first appears at order $\frac{\alpha_s^2}{m_Q^2}$ [13] [4], while in our model the leading spin-dependent effects are of order $\frac{\alpha_s}{m_Q^2}$. Since $\alpha_s$ is small in the heavy quark system ($\alpha_s(m_c^2) \sim 0.35$ and $\alpha_s(m_b^2) \sim 0.22$), we expect the annihilation effects on $Q\bar{Q}$ spectra to be small.

## III. SOLUTION OF THE WAVE EQUATIONS

### A. $Q\bar{q}$ sector

The Dirac equation (20) can be reduced by using the explicit forms of the zeroth order wave function (21) and the Dirac matrices $\boldsymbol{\alpha}$ and $\beta$ along with the identity

$$\boldsymbol{\sigma}^{(1)} \cdot \hat{\mathbf{r}} \mathcal{Y}_{\ell \frac{1}{2} j_1}^{m_{j_1}}(\Omega) = -\mathcal{Y}_{\bar{\ell} \frac{1}{2} j_1}^{m_{j_1}}(\Omega) \tag{52}$$

to extract the coupled radial wave equations [14]

$$\frac{dG_{n\ell j_1}(r)}{dr} + \frac{\kappa_1}{r} G_{n\ell j_1}(r) = (m_1 + V_s(r) - V_v(r) + E_{n\ell j_1}^{(0)}) F_{n\ell j_1}(r), \tag{53}$$

$$\frac{dF_{n\ell j_1}(r)}{dr} - \frac{\kappa_1}{r} F_{n\ell j_1}(r) = (m_1 + V_s(r) + V_v(r) - E_{n\ell j_1}^{(0)}) G_{\ell j_1}(r), \tag{54}$$

where

$$E_{n\ell j_1}^{(0)} = W_{n\kappa_1 j_1}^{(0)} - m_2. \tag{55}$$

We have obtained three separate numerical solutions of these coupled equations using two different techniques, direct integration and the matrix diagonalization-variational technique.

#### *1. Direct Integration*

This approach uses stepping techniques to obtain solutions to the differential equations. Such techniques are much more efficient if any large asymptotic damping of the radial wave functions can be extracted and reduced radial wave equations can then be integrated. The scale of the asymptotic variation of the radial wave functions is determined by the string tension $b$ appearing in the scalar potential (27). Defining a dimensionless radial variable $\rho = b^{1/2} r$, and determining the asymptotic behavior of the radial wave functions, the reduced wave functions $g(\rho)$ and $f(\rho)$ are defined in terms of $G$ and $F$ by

$$\begin{aligned} G(r) &= g(\rho) e^{-\frac{1}{2}(\rho^2 + \gamma\rho)}, \\ F(r) &= f(\rho) e^{-\frac{1}{2}(\rho^2 + \gamma\rho)}, \end{aligned} \tag{56}$$

where $\gamma = 2(m_1 + c)/b^{1/2}$, and $c$ is the constant shift in the scalar potential. Coupled equations for the reduced wave functions that result are

$$\left( \frac{d}{d\rho} - \rho - \frac{\gamma}{2} + \frac{\kappa_1}{\rho} \right) g(\rho) = \left( \alpha_+ + \rho - \overline{V}_v(\rho) \right) f(\rho), \tag{57}$$

$$\left( \frac{d}{d\rho} - \rho - \frac{\gamma}{2} - \frac{\kappa_1}{\rho} \right) f(\rho) = \left( \alpha_- + \rho + \overline{V}_v(\rho) \right) g(\rho), \tag{58}$$

where $\overline{V}_v(\rho) = V_v(r)/b^{1/2}$, $\alpha_\pm = \frac{\gamma}{2} \pm \varepsilon$ and $\varepsilon = E_{n\ell j_1}^{(0)}/b^{1/2}$.

In order to integrate the differential equations it is necessary to know the values of the functions and their derivatives at some point and then to have a stepping algorithm that predicts the values of the functions and their derivatives at subsequent points. The values of the functions and their first derivatives at $\rho = 0$ are obtained by construction of a series solution for the functions for small $\rho$. An adaptive Runge-Kutte routine [15] is used to integrate the differential equations for increasing values of $\rho$. Energy eigenvalues can be found by adjusting the value of the energy until the functions have the correct asymptotic behavior as determined by an asymptotic expansion of the functions at some large finite $\rho$. This process of finding the eigenenergies is called the shooting method [15]. In the calculations shown here, the accuracy of the eigenvalues is increased by integrating up from $\rho = 0$ and down from some large finite $\rho$ to some intermediate point where the values of $g(\rho)$ and $f(\rho)$ are required to match.

A second variation on this method is to use the reduced radial wave equations (57) and (58) to eliminate $f(\rho)$ to obtain a second order differential equation for $g(\rho)$. This equation can then be integrated in a manner similar to the Shrödinger equation for the $Q\bar{Q}$ sector.

### 2. Variational Method

The starting point for the 'variational' solution of eqs. (53, 54) is the pair of equations

$$
\begin{aligned}
E^{(0)}_{n\ell j}\mathcal{G}^n_{j\ell}(r) &= (m_1 + V_s + V_v)\mathcal{G}^n_{j\ell}(r) + \frac{\kappa_1 - 1}{r}\mathcal{F}^n_{j\bar{\ell}}(r) - \frac{d\mathcal{F}^n_{j\bar{\ell}}(r)}{dr},\\
E^{(0)}_{n\ell j}\mathcal{F}^n_{j\bar{\ell}}(r) &= (V_v - m_1 - V_s)\mathcal{F}^n_{j\bar{\ell}}(r) + \frac{\kappa_1 + 1}{r}\mathcal{G}^n_{j\ell}(r) + \frac{d\mathcal{G}^n_{j\ell}(r)}{dr},\\
\mathcal{F}^n_{j\bar{\ell}}(r) &= \frac{F_{n\ell j}(r)}{r},\ \ \mathcal{G}^n_{j,\ell}(r) = \frac{G_{n\ell j}(r)}{r}.
\end{aligned} \tag{59}
$$

The functions $\mathcal{F}$ and $\mathcal{G}$ are expanded in a set of orthonormal basis functions $\phi^i_\ell(r/\varrho)$

$$
\begin{aligned}
\mathcal{G}^n_{j\ell}(r) &= \sum_{i=1}^N \alpha^n_i \phi^i_\ell(r/\varrho),\\
\mathcal{F}^n_{j\bar{\ell}}(r) &= \sum_{i=1}^N \beta^n_i \phi^i_{\bar{\ell}}(r/\varrho),
\end{aligned} \tag{60}
$$

with

$$
\int_0^\infty drr^2\phi^{i*}_\ell(r/\varrho)\phi^k_\ell(r/\varrho) = \delta_{i,k}. \tag{61}
$$

$\varrho$ is the size parameter of the wave functions, and is used as the variational parameter in this calculation.

Substituting the expansion of eq. (60) into eq. (59), multiplying by $\phi^{k*}_{\ell(\bar{\ell})}(r/\varrho)$ and integrating, leads to the set of equations

$$
\begin{aligned}
E^{(0)}_{n\ell j}\alpha^n_k &= \sum_{i=1}^N \Big\langle m_1 + V_s(r) + V_v(r)\Big\rangle_{k\ell,i\ell}\alpha^n_i\\
&+ \sum_{i=1}^N \Big\langle \frac{\kappa_1 - 1}{r}\Big\rangle_{k\ell,i\bar{\ell}}\beta^n_i - \sum_{i=1}^N \Big\langle \frac{d}{dr}\Big\rangle_{k\ell,i\bar{\ell}}\beta^n_i,\\
E^{(0)}_{n\ell j}\beta^n_k &= \sum_{i=1}^N \Big\langle V_v(r) - m_1 - V_s(r)\Big\rangle_{k\bar{\ell},i\bar{\ell}}\beta^n_i\\
&+ \sum_{i=1}^N \Big\langle \frac{\kappa_1 + 1}{r}\Big\rangle_{k\bar{\ell},i\ell}\alpha^n_i + \sum_{i=1}^N \Big\langle \frac{d}{dr}\Big\rangle_{k\bar{\ell},i\ell}\alpha^n_i,
\end{aligned} \tag{62}
$$

where we use the symbolic notation

$$
\Big\langle \psi(r)\Big\rangle_{k\ell_1,i\ell_2} = \int_0^\infty drr^2\phi^{k*}_{\ell_1}(r/\varrho)\psi(r)\phi^i_{\ell_2}(r/\varrho). \tag{63}
$$

The two sets of equations represented by eq. (62) can be combined into the single eigenvalue equation

$$
\begin{pmatrix} \langle m + V_s(r) + V_v(r) - E\rangle & \langle \frac{\kappa_1 - 1}{r} - \frac{d}{dr}\rangle \\ \langle \frac{\kappa_1+1}{r} + \frac{d}{dr}\rangle & \langle V_v(r) - m - V_s(r) - E\rangle \end{pmatrix}\begin{pmatrix}\alpha\\ \beta\end{pmatrix} = 0. \tag{64}
$$

The size of the matrix in eq. (64) is $2N \times 2N$. Solutions to the eq. (59) are obtained by varying the wave function size parameter $\varrho$, diagonalising the matrix in eq. (64) for each value of $\varrho$, and searching for stationary points in the eigenvalues as functions of $\varrho$. In principle, if the size of the expansion basis $N$ is taken to $\infty$, solutions obtained in this way would be exact and independent of $\varrho$. In practice, the procedure outlined above is carried out for finite $N$, and $N$ is increased until the eigenvalues are largely independent of $\varrho$, for some reasonable range in $\varrho$. With this method, the lower $N$ eigenvalues obtained correspond to negative energy states, while the higher $N$ eigenvalues are those of interest for this problem.

For this problem we have used harmonic oscillator wave functions for the expansion, with $N = 10$ and $N = 20$. We compare the numerical solutions that we obtain using this procedure with those that are obtained using the other previously described methods. As expected, the variational solutions are better for $N = 20$, and the eigenvalues are within 1% of those obtained by solving the equations by the methods described in the previous subsection.

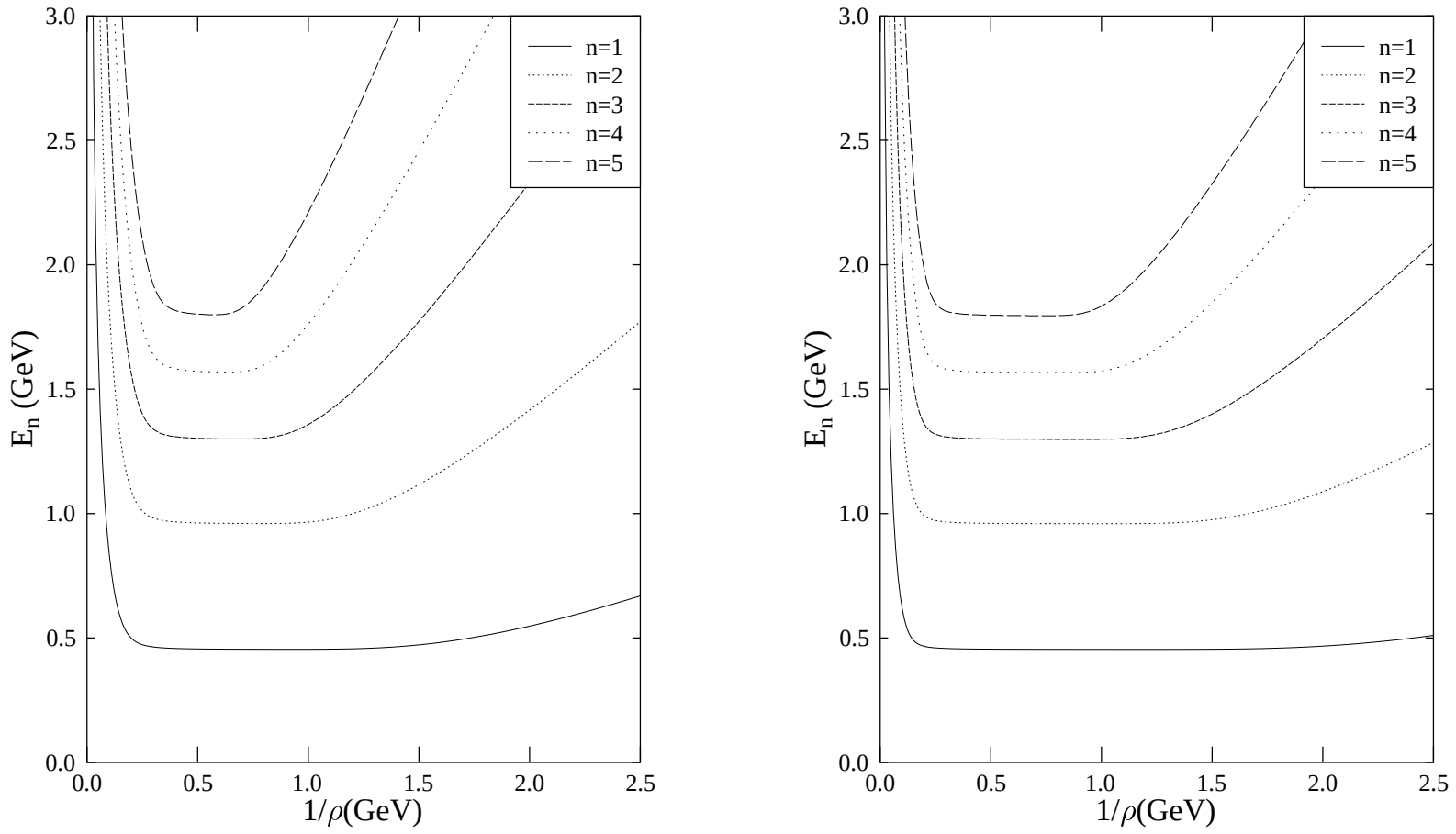

FIG. 2. Energy eigenvalues as a function of $\frac{1}{\varrho}$, for $N = 10$ and $N = 20$ .

### B. $Q\bar{Q}$ sector

Using eq. (48) in eq. (47) and defining $\rho = b^{1/2} r$, the differential equation for the radial wave function is

$$\left[ -\frac{1}{\mu} \left( \frac{d^2}{d\rho^2} - \frac{L(L+1)}{\rho^2} \right) + \overline{V}_v(\rho) + \rho \right] u_{nL}(\rho) = \varepsilon u_{nL}(\rho), \tag{65}$$

where $\mu = m_Q/b^{1/2}$ , $\varepsilon = (W_{nL}^{(0)} - 2m_Q - c)/b^{1/2}$ and $\overline{V}_v(\rho) = V_v(r)/b^{1/2}$. Determining the asymptotic behavior of the radial wave function, the reduced radial wave function $g(\rho)$ can be defined by

$$u_{nL}(\rho) = g(\rho) e^{-\mu^{\frac{1}{2}}(\frac{2}{3}\rho^{\frac{3}{2}} - \varepsilon \rho^{\frac{1}{2}})} \tag{66}$$

The appearance of fractional powers of $\rho$ in the argument of the exponential function in (66) leads to coefficients with fractional powers of $\rho$ in the differential equation for $g(\rho)$. This complicates the expansion of the reduced radial wave functions for small and large values of $\rho$. It is, therefore, convenient to define the variable $\xi = \rho^{1/2}$. The differential equation for $g(\xi)$ can then be written as

$$\left[ -\xi^2 \frac{d^2}{d\xi^2} + \left( \xi - 2\varepsilon\mu^{\frac{1}{2}}\xi^2 + 4\mu\xi^4 \right) \frac{d}{d\xi} + \left( 4L(L+1) + \mu^{\frac{1}{2}}(\varepsilon\xi + 2\xi^3) - \mu\varepsilon^2\xi^2 + 4\mu\xi^4 \overline{V}_v(\xi^2) \right) \right] g(\xi) = 0 \tag{67}$$

This equation can be used to develop expansions for small and large $\xi$ to provide boundary conditions for numerical integration of the differential equation.

Since the Runge-Kutte method is designed to integrate systems of coupled first-order differential equations it is necessary to reexpress the differential equation (67) as the coupled pair

$$\frac{d}{d\xi} g(\xi) = f(\xi), \tag{68}$$

and

$$\left[ -\xi^2 \frac{d}{d\xi} + \left( \xi - 2\varepsilon\mu^{\frac{1}{2}}\xi^2 + 4\mu\xi^4 \right) \right] f(\xi) + \left( 4L(L+1) + \mu^{\frac{1}{2}}(\varepsilon\xi + 2\xi^3) - \mu\varepsilon^2\xi^2 + 4\mu\xi^4 \overline{V}_v(\xi^2) \right) g(\xi) = 0. \tag{69}$$

This system can then be solved by Runge-Kutte integration and intermediate-point shooting techniques.

## IV. RESULTS

Once the zeroth-order solutions are found, the perturbed energies can be calculated using (24) and (50). The masses associated with the bound states are given by (26) and (51). These depend on the quark masses $m_u$, $m_s$, $m_c$ and $m_b$ as applicable for each meson; the parameters of the scalar potential (27) $b$ and $c$; and the parameters of the vector potential (28) $\alpha_i$ and $\gamma_i$ for $i = 1, 2, 3$. The model contains a total of twelve parameters. In obtaining the results shown here, the vector potential parameters

$$\alpha_2 = 0.15, \qquad \alpha_3 = 0.2,$$
$$\gamma_1 = 0.5, \qquad \gamma_2 = 1.581, \qquad \gamma_3 = 15.81, \tag{70}$$

are fixed at the same values as given in ref. [4]. The remaining vector potential parameter $\alpha_1$ is reexpressed as

$$\alpha_1 = \alpha_{crit} - \alpha_2 - \alpha_3. \tag{71}$$

TABLE II. Parameters of the model.

| parameter | value | comments |
|---|---|---|
| $\alpha_{crit}$ | 0.674 | limiting value of $\alpha_s$ |
| $b$ | 0.180 GeV$^2$ | string tension |
| $c$ | 0.02 GeV | see eq. (27) |
| $m_u$ | 0.258 GeV | |
| $m_s$ | 0.400 GeV | |
| $m_c$ | 1.53 GeV | |
| $m_b$ | 4.87 GeV | |

TABLE III. Fitted meson spectra for $Q\bar{q}$ mesons.

| | | Mass (GeV) | |
|---|---|---|---|
| Meson | $J^P$ | theory | experiment[a] |
| $D$ | $0^-$ | 1.85 | 1.87 |
| $D^*$ | $1^-$ | 2.02 | 2.01 |
| $D_1$ | $1^+$ | 2.41 | 2.42 |
| $D_2^*$ | $2^+$ | 2.46 | 2.46 |
| $B$ | $0^-$ | 5.28 | 5.28 |
| $B^*$ | $1^-$ | 5.33 | 5.33 |
| $D_s$ | $0^-$ | 1.94 | 1.97 |
| $D_s^*$ | $1^-$ | 2.13 | 2.11 |
| $B_s$ | $0^-$ | 5.37 | 5.38 |
| $B_s^*$ | $1^-$ | 5.43 | 5.43 |

[a] Experimental values are quoted [16] to the nearest 10 MeV due to ambiguities in assigning the calculated values to specific charge states.

where $\alpha_{crit}$ is the value of the running coupling constant at $Q^2 = 0$ as parametrized in ref. [4].

$\alpha_{crit}$ and the remaining model parameters are adjusted to fit the masses of a selection of mesons. The resulting values are listed in Table II. The fitted meson spectra for the $Q\bar{q}$ sector are listed in Table III and the fitted meson spectra for the $Q\bar{Q}$ are listed in Table IV. Additional states which were not used in the fitting procedure were calculated and a detailed discussion of the results for the $Q\bar{q}$ and $Q\bar{Q}$ is presented in the following two subsections.

TABLE IV. Fitted meson spectra for $Q\bar{Q}$ mesons.

| | | Mass (GeV) | |
|---|---|---|---|
| Meson | $J^{PC}$ | theory | experiment |
| $\eta_c$ | $0^{-+}$ | 3.00 | 2.98 |
| $J/\psi(1S)$ | $1^{--}$ | 3.10 | 3.10 |
| $\chi_{c0}$ | $0^{++}$ | 3.44 | 3.42 |
| $\chi_{c1}$ | $1^{++}$ | 3.50 | 3.51 |
| $\chi_{c2}$ | $2^{++}$ | 3.54 | 3.56 |
| $J/\psi(2S)$ | $1^{--}$ | 3.73 | 3.69 |
| $\Upsilon(1S)$ | $1^{--}$ | 9.46 | 9.46 |
| $\chi_{b0}(1P)$ | $0^{++}$ | 9.85 | 9.86 |
| $\chi_{b1}(1P)$ | $1^{++}$ | 9.87 | 9.89 |
| $\chi_{b2}(1P)$ | $2^{++}$ | 9.89 | 9.92 |
| $\Upsilon(2S)$ | $1^{--}$ | 10.02 | 10.02 |
| $\chi_{b0}(2P)$ | $0^{++}$ | 10.24 | 10.24 |
| $\chi_{b1}(2P)$ | $1^{++}$ | 10.26 | 10.26 |
| $\chi_{b2}(2P)$ | $2^{++}$ | 10.28 | 10.27 |
| $\Upsilon(3S)$ | $1^{--}$ | 10.39 | 10.36 |

### A. $Q\bar{q}$ sector

For the $Q\bar{q}$ sector, the zeroth-order eigenenergy $E^{(0)}_{n\ell j_1} = W^{(0)}_{n\kappa_1 j_1} - m_2$ is independent of the heavy quark mass, as would be expected in the heavy quark limit, where the heavy quark should act as a static source. The zeroth-order spectrum depends only on the light quark mass. The first-order correction to the mass $W^{(1)}_{nJP}$ is proportional to $1/m_2$ and splits each of the unperturbed states. These features are illustrated in Fig. 3 which shows $W^{(0)}_{n\kappa_1 j_1} - m_2$ for a $\bar{u}$ quark as solid lines and $W_{nJP} - m_2 = W^{(0)}_{n\kappa_1 j_1} + W^{(1)}_{nJP} - m_2$ with a $c$ quark as the heavy quark (dotdashed lines) and with a $b$ quark as the heavy quark (dashed lines). Fig. 4 is a similar spectrum where the light quark is now an $\bar{s}$ quark.

Note that to zeroth order the ordering of the $j_1 = \ell \pm 1/2$ states is reversed for the $\ell = 2$ states in comparison to the $\ell = 1$ states. This phenomenon, called multiplet inversion, has been predicted [17] for $Q\bar{q}$ mesons with $m_2 \gg m_1$. It results from the dominance of the Thomas-precession over the spin-dependent forces in this limit.

For the states presented here, the root mean square momentum of the zeroth-order wave function is approximately 0.9 GeV. Clearly, both $u$ and $s$ quarks are very relativistic. In addition, it is possible to obtain some sense of the convergence of the $p/m$ expansion for the corrections to the infinite-heavy-quark-mass limit since $\frac{p_{rms}}{m_c} \sim \frac{1}{2}$ while $\frac{p_{rms}}{m_b} \sim \frac{1}{5}$. Therefore, the higher-order correction that are neglected here should be considerably larger for the the $c$ quark than the $b$ quark. Indeed, this problem will become worse with increasing $n$ since $p_{rms}$ should increase with increasing $n$. This is seen in the shift of the $0^-$ states relative to the unperturbed states which increases with $n$.

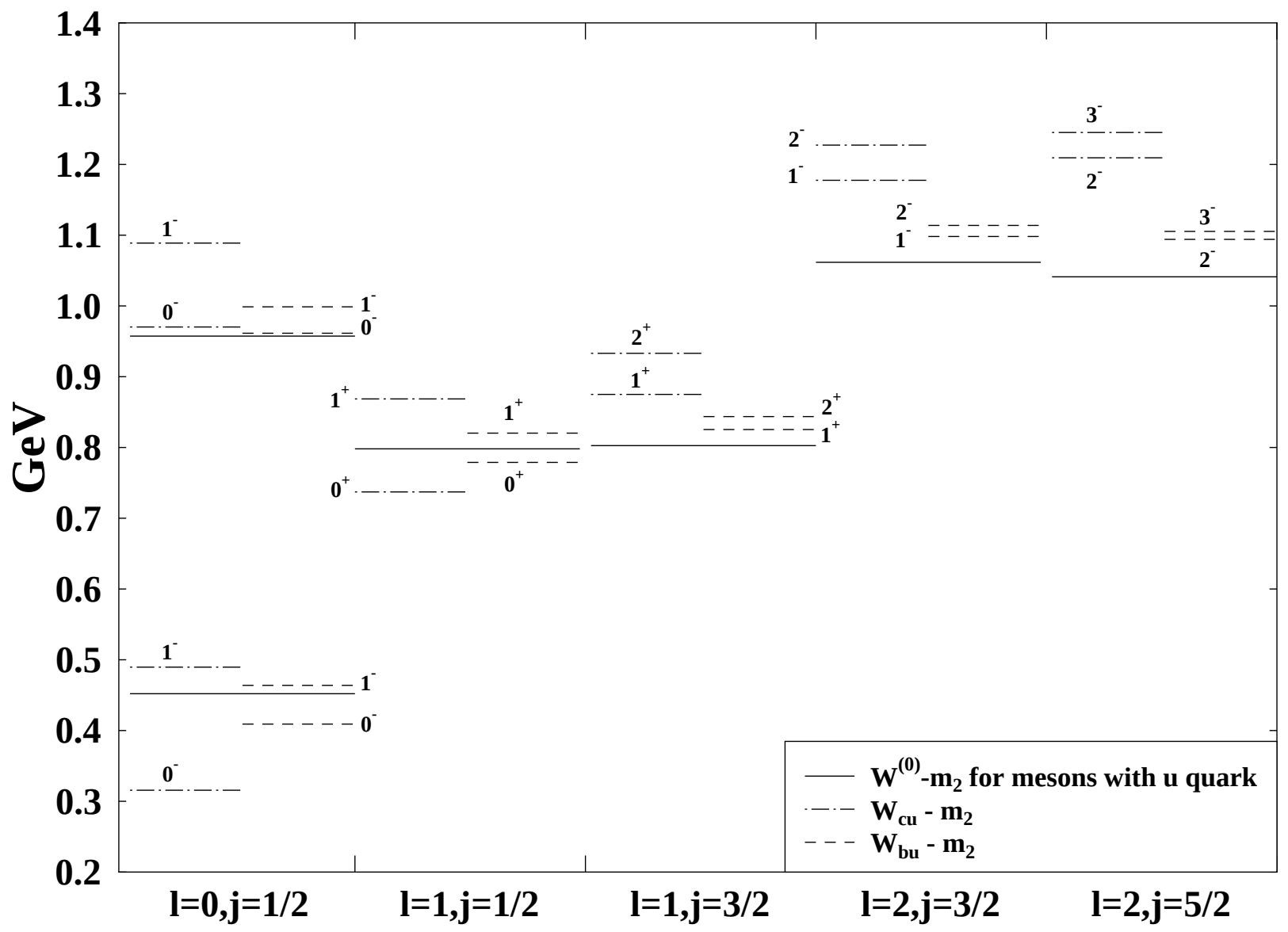

FIG. 3. This figure shows $W - m_2$ for $b\bar{u}$ and $c\bar{u}$ to the zeroth order and to the first order. $l_1$ and $j_1$ are the quantum numbers for orbital angular momentum and total angular momentum of the $\bar{u}$ quark. The states have been labelled as $J^P$.

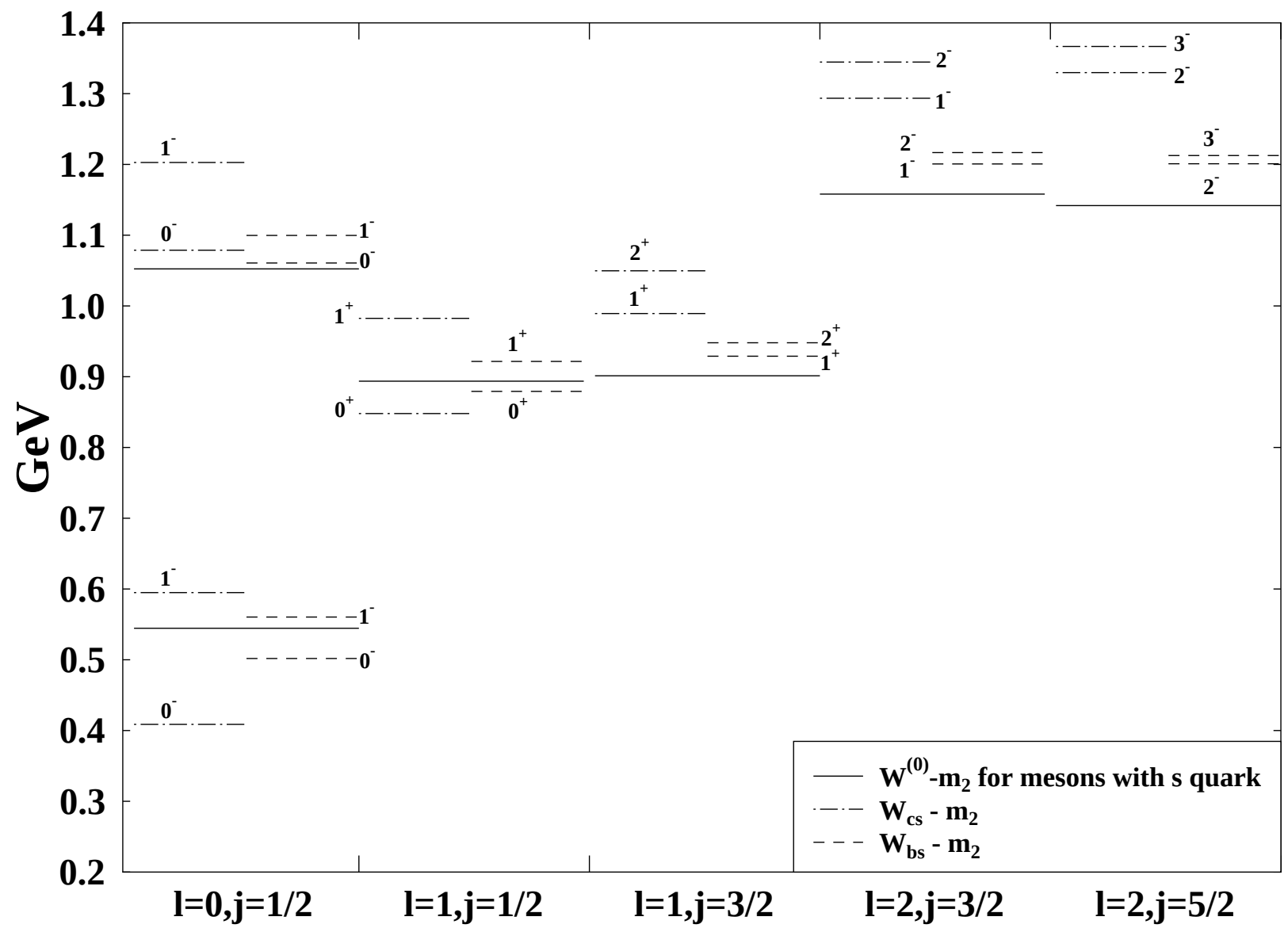

FIG. 4. This figure shows $W - m_2$ for $b\bar{s}$ and $c\bar{s}$ to the zeroth order and to the first order. $l_1$ and $j_1$ are the quantum numbers for orbital angular momentum and total angular momentum of the $\bar{u}$ quark. The states have been labelled as $J^P$.

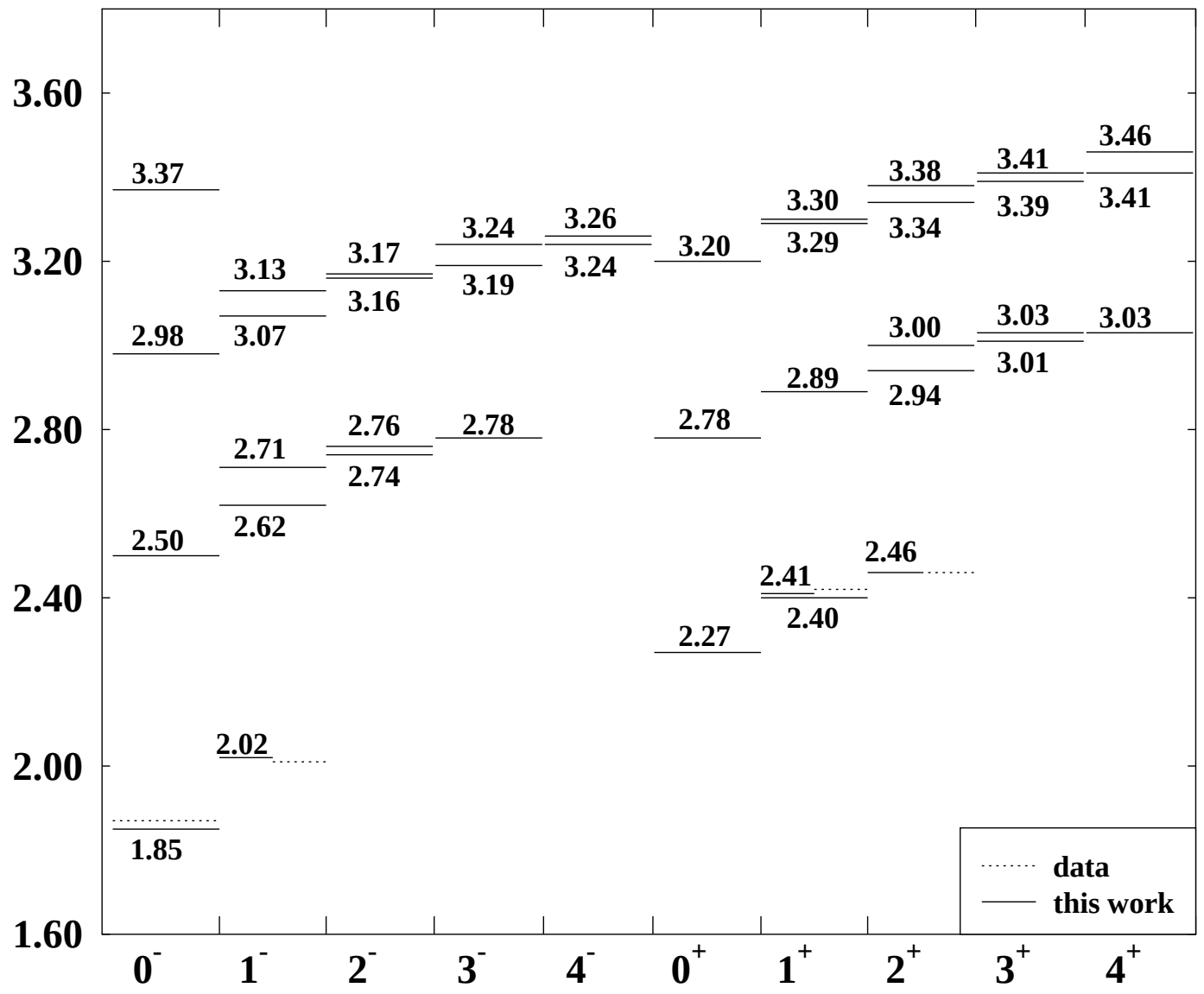

FIG. 5. $c\bar{u}$ spectrum. In this figure, solid lines represent the results of our calculation for the masses of $c\bar{u}$ mesons, $W$, to the first order in the perturbation; dotted lines represent the data.

Figs. 5 to 9 show predictions for the masses of $Q\bar{q}$ mesons, $W$, to first order in the perturbation (solid lines). In the spectra for mesons with $\bar{u}$ and $\bar{s}$ quarks, the available data are plotted for comparison as dotted lines. Ref. [16] has also listed states $D_J(2.440)$ and $D_{sJ}(2.573)$ with uncertain quantum numbers. We believe they correspond to the state $1^+(2.41)$ in Fig. 5 and the state $2^+(2.58)$ in Fig. 6 respectively. For the $b\bar{c}$ mesons, calculated masses from [4] are plotted because no data exist at present. For the $b\bar{c}$ mesons, $\frac{p_{rms}}{m_c} \sim 1$. This shows that although the mass of the $c$ quark is relatively large it is quite relativistic in this case.

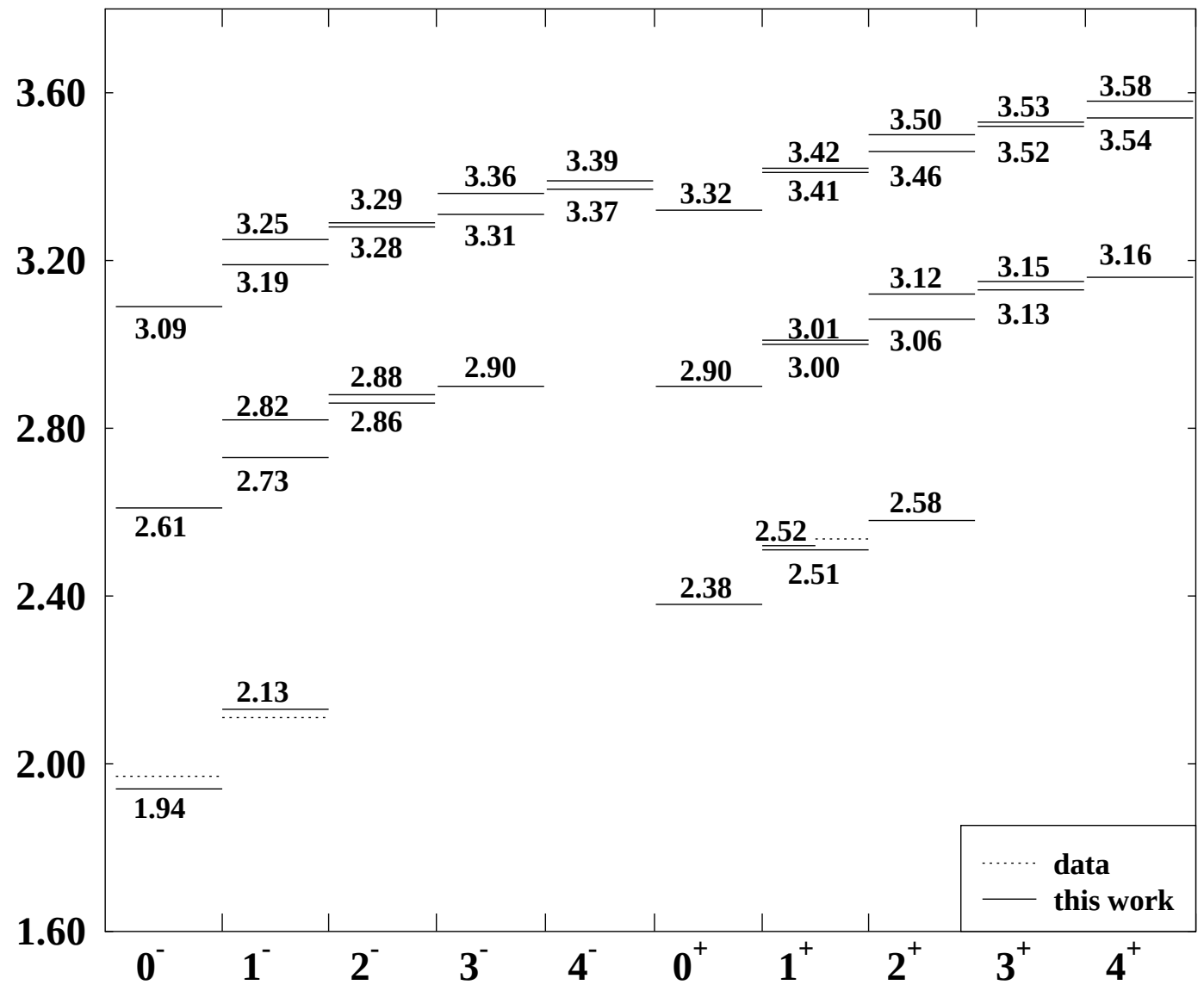

FIG. 6. $c\bar{s}$ spectrum. See caption of Fig. 5.

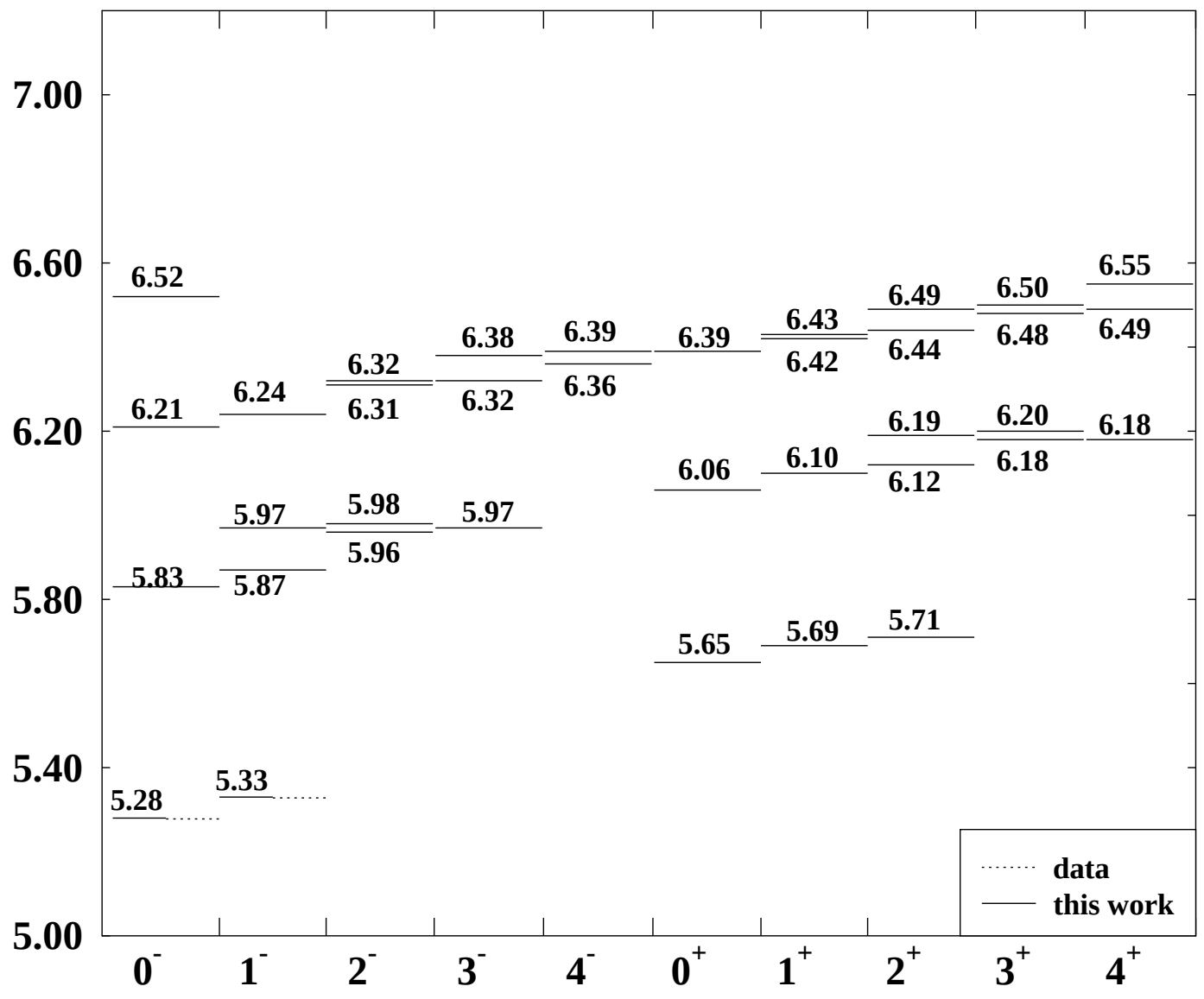

FIG. 7. $b\bar{u}$ spectrum. See caption of Fig. 5.

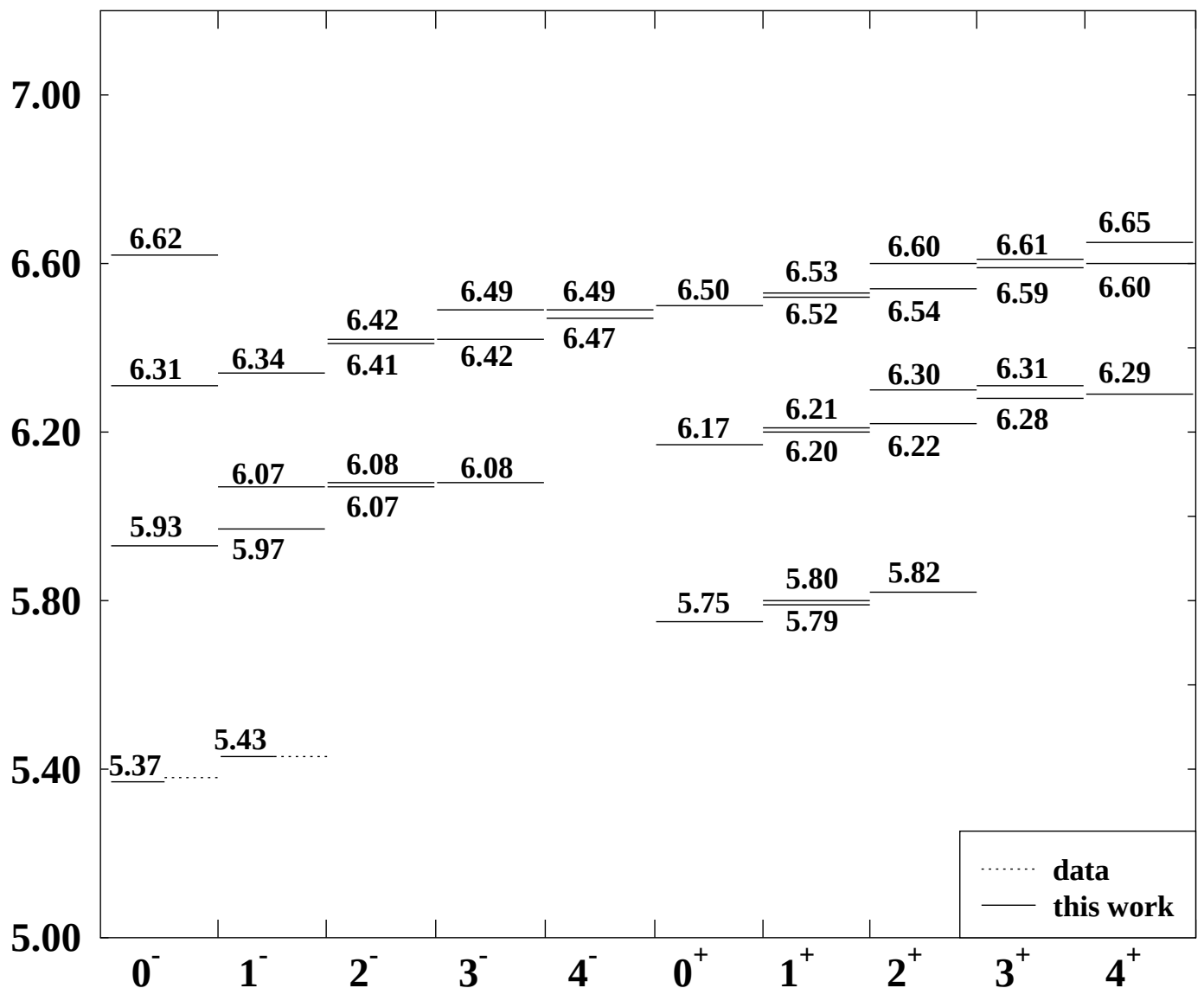

FIG. 8. $b\bar{s}$ spectrum. See caption of Fig. 5.

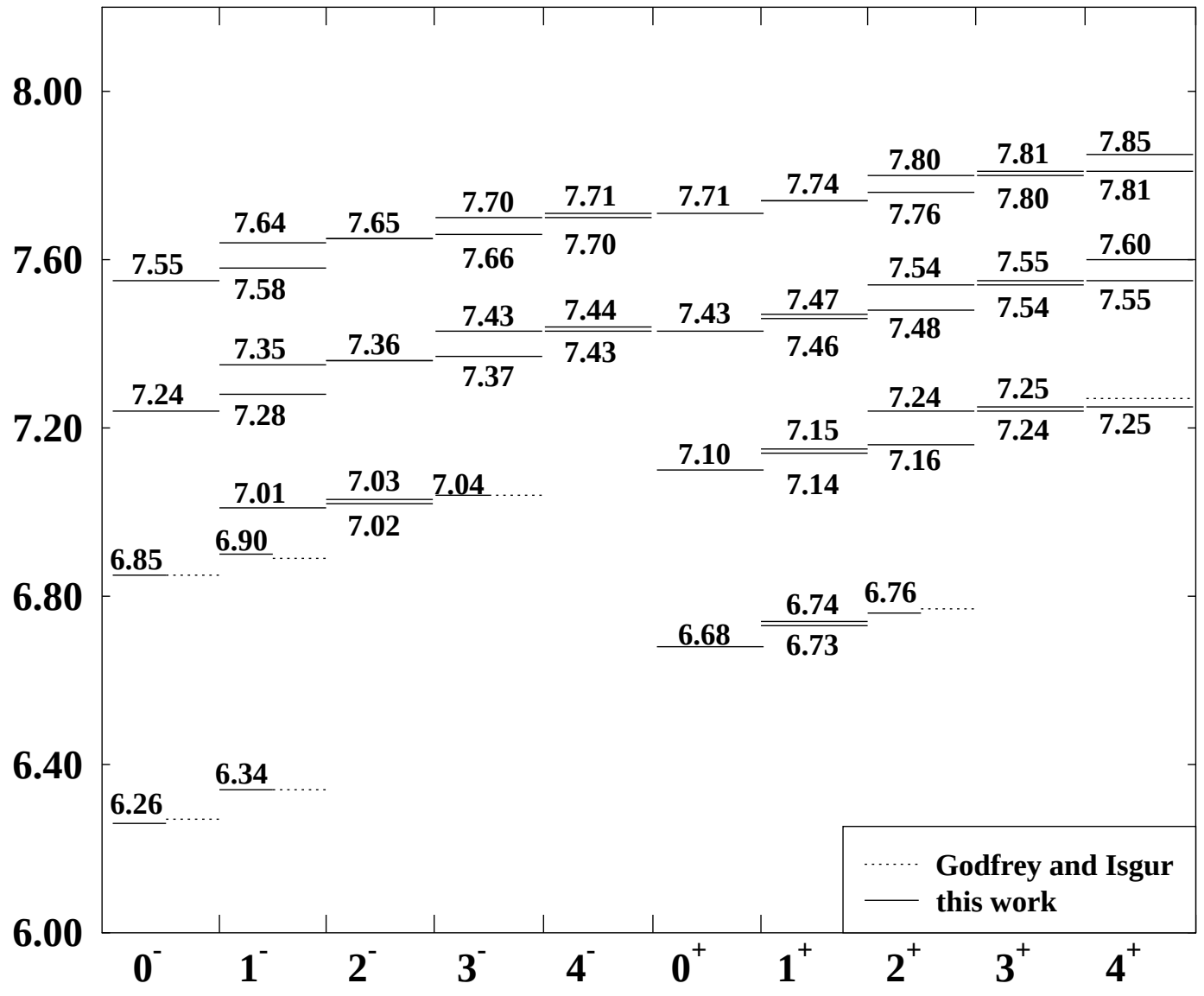

FIG. 9. $b\bar{c}$ spectrum. See caption of Fig. 5.

In these figures, the results are in good agreement with the data, which vindicates our choices of potentials and parameters. However, the calculated hyperfine splittings are all larger than in the data. The agreement is much better in the $b$-flavored mesons than in the $c$-flavored mesons. There are three possible reasons for this discrepancy. First, as has been mentioned earlier, this model is expected to work better for $b$-flavored mesons than for $c$-flavored mesons due to the more rapid convergence of the nonrelativistic expansion applied to the heavy quark. Secondly, these calculations do not include any effects associated with possible strong decay of the heavy mesons. The coupling to these strong decay channels will result in shifts in the meson masses as well as decay widths for heavy mesons above decay thresholds. These shifts will be greatest near the decay thresholds.

The third possible reason for the large hyperfine splittings may have its origin in the parametrization of $\alpha_s(r)$, particularly at small $r$. While many functional forms may be used for this parametrization, each form may be expected to lead to quite different $1/m_Q$ contributions, especially in the hyperfine term. This question is currently under investigation.

The third term on the right hand side of (18b) has off-diagonal matrix elements between states with $j_1$ differing by unity and with $\ell$ differing by either 0 or 2. These mixings do not affect the spectrum to order $\frac{1}{m_Q}$ but should result in shifts in some states at higher order in all of these systems. This should be particularly apparent for the $1^+$ states which are nearly degenerate to order $\frac{1}{m_Q}$ for all $Q\bar{q}$ mesons calculated here.

One very interesting aspect of this calculation is the mapping of our model onto the heavy quark effective theory, with a view to evaluating some of the parameters and dynamical quantities (such as universal form factors) of the effective theory. While we do not endeavor to perform such a calculation for all such quantities here, some comments are merited.

Although we have included all of the $1/m_Q$ terms that arise from the spectator equation, it is not clear that these correspond to all of the $1/m_Q$ terms of HQET. In particular, in the spectator equation, the heavy quark is treated as being *exactly* on its mass shell. In contrast, in HQET, the heavy quark is allowed to be slightly off its mass shell (via the equation $p_\mu = m_Q v_\mu + k_\mu$), and this leads to terms that may be absent from the formulation presented here. The full ramifications of this are also under investigation.

Until this question is resolved, we dare not examine quantities that are intimately bound up in the $1/m_Q$ structure of the effective theory or the model. We can, however, examine quantities that depend only on the leading-order structure of the model, as we believe that this is a reasonably accurate representation of the effective theory. In particular, in the effective theory, one expects that the heavy quark should act as a static color source. This very important feature is reproduced in the model, as the leading dynamical behavior is described in terms of a Dirac equation for the light quark.

Two quantities of interest in HQET are $\bar{\Lambda}$ and $\lambda_1$, which are defined by

$$M_M = m_Q + \bar{\Lambda} + \mathcal{O}\left(\frac{1}{m_Q}\right),$$

$$\left\langle M(v)\left|\bar{h}_Q(iD)^2 h_Q\right| M(v)\right\rangle = 2M_M\lambda_1.$$

$\bar{\Lambda}$ is crucial for the effective theory, as it appears as the coefficient in the $1/m_Q$ expansion: the expansion coefficient is written as $\bar{\Lambda}/m_Q$. $\bar{\Lambda}$ is, in essence, the contribution to the mass of the meson from the mass and kinetic energy of the "brown muck". The left hand side of the second expression above is proportional to the kinetic energy of the heavy quark. The meson states in the bra and ket above are the leading order representation, and so correspond to our zeroth-order calculation. From our model, we obtain $\bar{\Lambda} = 0.45$ GeV for the ground state pseudoscalar/vector doublet, and $\lambda_1 = 0.67$ GeV$^2$. These values are in reasonable agreement with other values in the literature [3]. Further aspects of the relationship of our model to HQET are discussed in the conclusions.

### B. $Q\bar{Q}$ sector

Figs. 10 and 11 show the spectra for $c\bar{c}$ and $b\bar{b}$ mesons as calculated with eqs. (44)-(51). As before, the calculated masses are shown as solid lines and the experimental masses as dotted lines. The $D\bar{D}$ and $B\bar{B}$ thresholds are shown as horizontal dotdashed lines across the Figs. 10 and 11 respectively. Ref. [16] has also listed states $h_c(1P)$ with mass 3.526 GeV and $\eta_c(2S)$ with mass 3.590 GeV. We believe they correspond to the states $2^1S_0(3.67)$ and $1^1P_1(3.51)$ in Fig. 10 respectively.

The $b\bar{b}$ spectrum is in quite good agreement with the data for the states lying below the $BB$ threshold. The agreement deteriorates as the masses approach and cross the $BB$ threshold. As argued in the previous section, this may be the result of the absence of coupling to strong decay channels. The agreement for the $c\bar{c}$ is less satisfactory. This may be an indication of the inadequacy of the truncation of the nonrelativistic expansion at order $\frac{1}{m_Q^2}$. In both cases the hyperfine splitting of the spin triplet states is too large.

Since the hyperfine tensor interaction has non-zero off diagonal matrix elements for states with spin 1 and with $L$ differing by 0 or 2, there should be mixings of states such as $^3S_1$ with $^3D_1$ and $^3P_2$ with $^3F_2$. These mixings do not affect the spectrum to order $\frac{1}{m_Q^2}$ but should result in shifts in some states at higher order in both the $b\bar{b}$ and $c\bar{c}$ spectra.

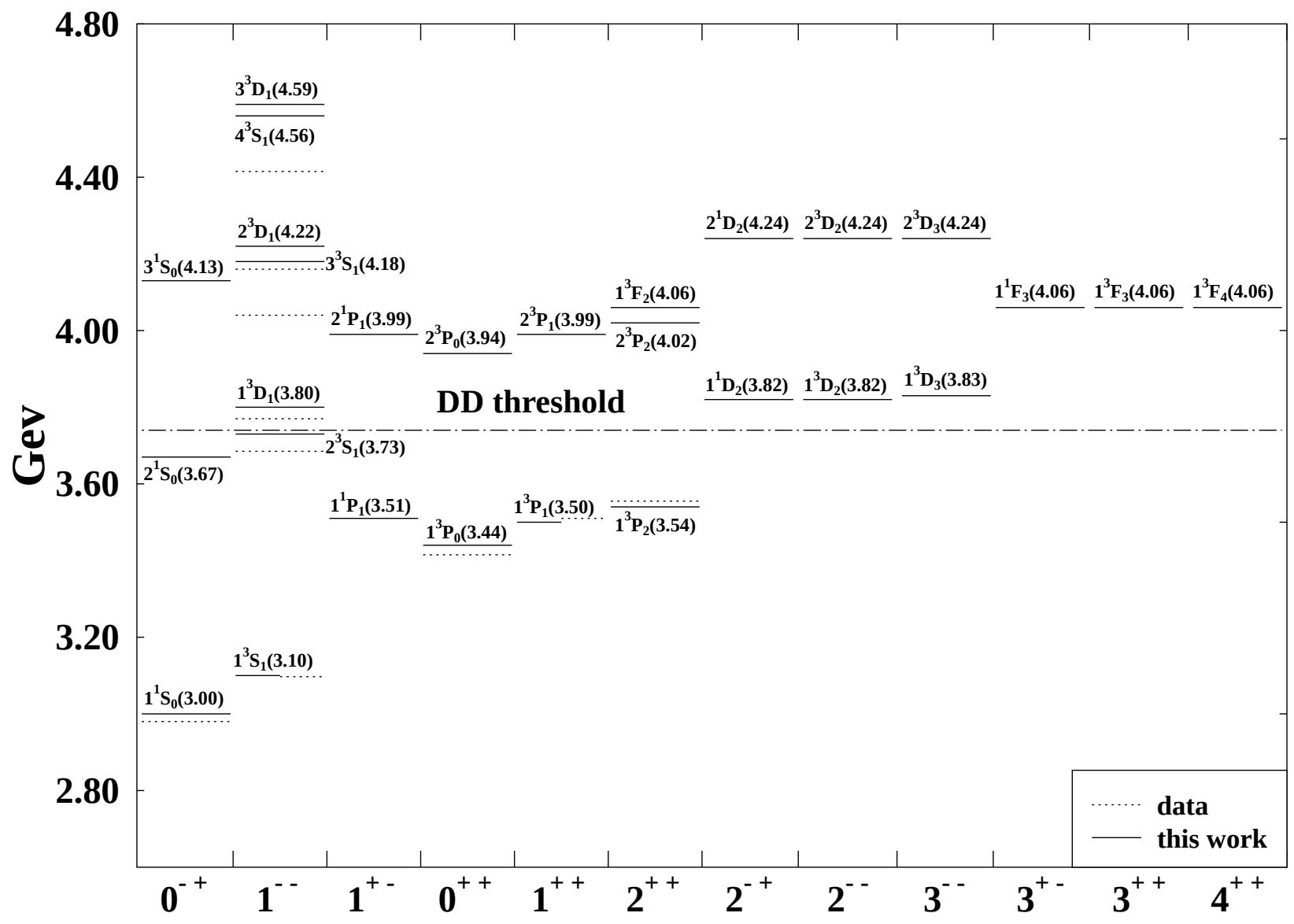

FIG. 10. $c\bar{c}$ spectra. See caption of Fig. 5.

Table V shows the individual contributions to the masses $W$ of a number of $b\bar{b}$ states from $W^{(0)}$, $E_{\rm c}$, $E_{\rm hyp}$, $E_{\rm so}$, $E_{\rm SR}$ and $E_{\rm VR}$. The retardation contributions $E_{\rm SR}$ and $E_{\rm VR}$ are clearly gauge dependent since they would not appear in the Coulomb gauge. $E_{\rm c}$ is also gauge dependent. These contributions may also be sensitive to the choice of quasipotential prescription. To this order $E_{\rm hyp}$, $E_{\rm so}$ should be independent of these factors. Note that the scalar and vector retardation contributions are of opposite sign and therefore tend to cancel. However the sum of these contributions is comparable with $E_{\rm hyp}$ and $E_{\rm so}$. The assumption that the scalar retardation potential depends only on the square of the exchanged four-momentum $Q^2$ is uncontrolled and it is possible to propose forms for this retardation potential which would eliminate the scalar term altogether.

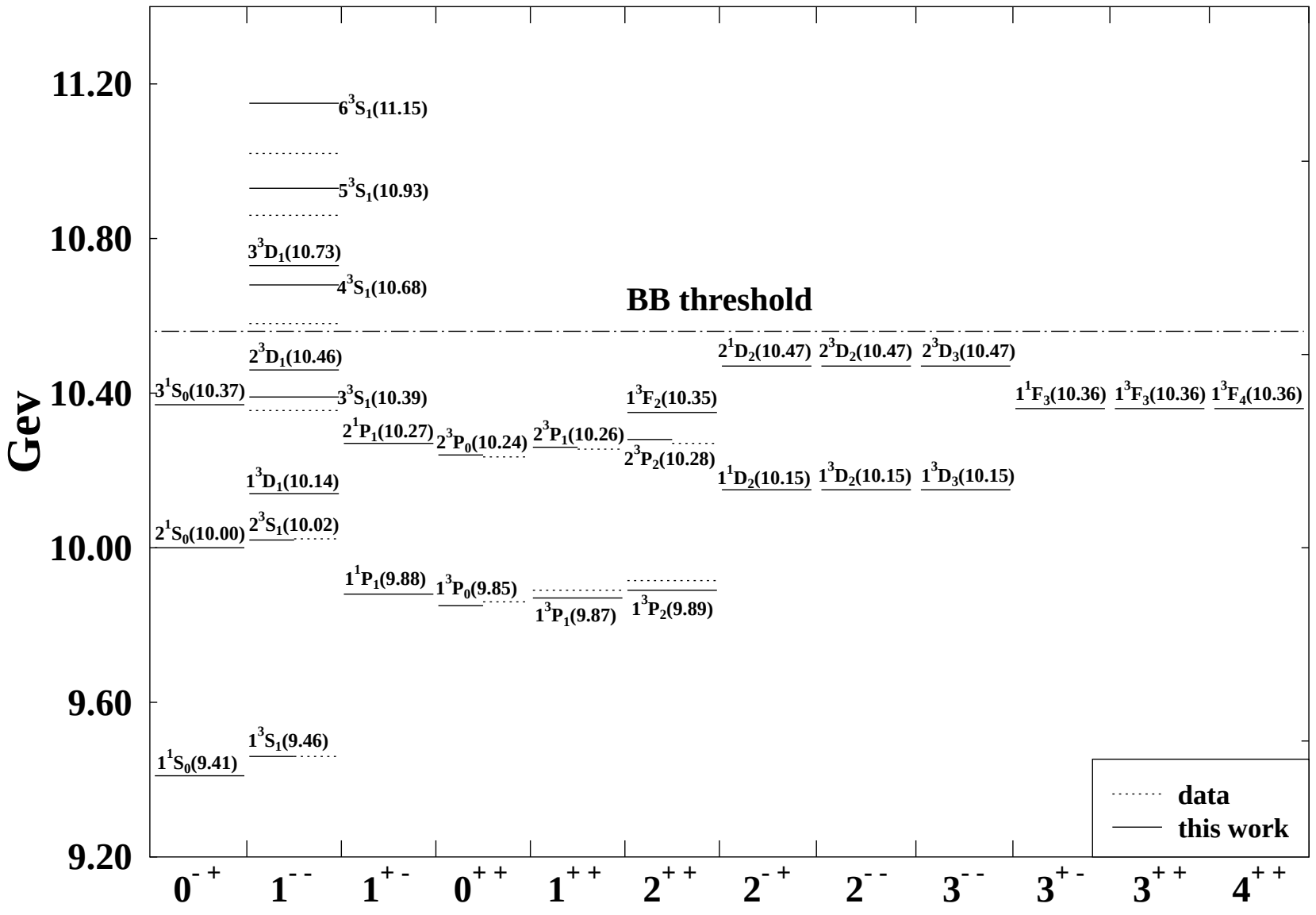

FIG. 11. $b\bar{b}$ spectra. See caption of Fig. 5.

TABLE V. Zeroth order and various first order interaction energies in the $b\bar{b}$ spectrum (GeV)

| State | $W$ | $W^{(0)}$ | $E_{\rm c}$ | $E_{\rm hyp}$ | $E_{\rm so}$ | $E_{\rm SR}$ | $E_{\rm VR}$ | $E_{\rm SR}+E_{\rm VR}$ |
|---|---|---|---|---|---|---|---|---|
| $1^1S_0$ | 9.41 | 9.5315 | -0.0602 | -0.0367 | 0.0000 | 0.0072 | -0.0297 | -0.0224 |
| $1^3S_1$ | 9.46 | 9.5315 | -0.0602 | 0.0122 | 0.0000 | 0.0072 | -0.0297 | -0.0224 |
| $2^1S_0$ | 10.00 | 10.0892 | -0.0708 | -0.0192 | 0.0000 | 0.0175 | -0.0192 | -0.0017 |
| $2^3S_1$ | 10.02 | 10.0892 | -0.0708 | 0.0064 | 0.0000 | 0.0175 | -0.0192 | -0.0017 |
| $3^1S_0$ | 10.37 | 10.4511 | -0.0839 | -0.0146 | 0.0000 | 0.0302 | -0.0160 | 0.0142 |
| $3^3S_1$ | 10.39 | 10.4511 | -0.0839 | 0.0049 | 0.0000 | 0.0302 | -0.0160 | 0.0142 |
| $4^1S_0$ | 10.66 | 10.7411 | -0.0992 | -0.0125 | 0.0000 | 0.0447 | -0.0144 | 0.0303 |
| $4^3S_1$ | 10.68 | 10.7411 | -0.0992 | 0.0042 | 0.0000 | 0.0447 | -0.0144 | 0.0303 |
| $5^1S_0$ | 10.91 | 10.9928 | -0.1162 | -0.0113 | 0.0000 | 0.0608 | -0.0135 | 0.0473 |
| $5^3S_1$ | 10.93 | 10.9928 | -0.1162 | 0.0038 | 0.0000 | 0.0608 | -0.0135 | 0.0473 |
| $6^1S_0$ | 11.14 | 11.2202 | -0.1345 | -0.0105 | 0.0000 | 0.0781 | -0.0128 | 0.0653 |
| $6^3S_1$ | 11.15 | 11.2202 | -0.1345 | 0.0035 | 0.0000 | 0.0781 | -0.0128 | 0.0653 |
| $1^1P_1$ | 9.88 | 9.9438 | -0.0610 | -0.0023 | 0.0000 | 0.0126 | -0.0169 | -0.0043 |
| $1^3P_0$ | 9.85 | 9.9438 | -0.0610 | -0.0074 | -0.0243 | 0.0126 | -0.0169 | -0.0043 |
| $1^3P_1$ | 9.87 | 9.9438 | -0.0610 | 0.0049 | -0.0121 | 0.0126 | -0.0169 | -0.0043 |
| $1^3P_2$ | 9.89 | 9.9438 | -0.0610 | -0.0001 | 0.0121 | 0.0126 | -0.0169 | -0.0043 |
| $2^1P_1$ | 10.27 | 10.3321 | -0.0752 | -0.0016 | 0.0000 | 0.0244 | -0.0143 | 0.0101 |
| $2^3P_0$ | 10.24 | 10.3321 | -0.0752 | -0.0056 | -0.0182 | 0.0244 | -0.0143 | 0.0101 |
| $2^3P_1$ | 10.26 | 10.3321 | -0.0752 | 0.0036 | -0.0091 | 0.0244 | -0.0143 | 0.0101 |
| $2^3P_2$ | 10.28 | 10.3321 | -0.0752 | -0.0001 | 0.0091 | 0.0244 | -0.0143 | 0.0101 |
| $1^1D_2$ | 10.15 | 10.2072 | -0.0637 | -0.0008 | 0.0000 | 0.0186 | -0.0139 | 0.0047 |
| $1^3D_1$ | 10.14 | 10.2072 | -0.0637 | -0.0011 | -0.0097 | 0.0186 | -0.0139 | 0.0047 |
| $1^3D_2$ | 10.15 | 10.2072 | -0.0637 | 0.0016 | -0.0032 | 0.0186 | -0.0139 | 0.0047 |
| $1^3D_3$ | 10.15 | 10.2072 | -0.0637 | -0.0001 | 0.0064 | 0.0186 | -0.0139 | 0.0047 |
| $2^1D_2$ | 10.47 | 10.5277 | -0.0792 | -0.0006 | 0.0000 | 0.0315 | -0.0125 | 0.0190 |
| $2^3D_1$ | 10.46 | 10.5277 | -0.0792 | -0.0009 | -0.0080 | 0.0315 | -0.0125 | 0.0190 |
| $2^3D_2$ | 10.47 | 10.5277 | -0.0792 | 0.0013 | -0.0027 | 0.0315 | -0.0125 | 0.0190 |
| $2^3D_3$ | 10.47 | 10.5277 | -0.0792 | -0.0001 | 0.0053 | 0.0315 | -0.0125 | 0.0190 |
| $1^1F_3$ | 10.36 | 10.4164 | -0.0717 | -0.0004 | 0.0000 | 0.0250 | -0.0124 | 0.0126 |
| $1^3F_2$ | 10.35 | 10.4164 | -0.0717 | -0.0004 | -0.0047 | 0.0250 | -0.0124 | 0.0126 |
| $1^3F_3$ | 10.36 | 10.4164 | -0.0717 | 0.0008 | -0.0012 | 0.0250 | -0.0124 | 0.0126 |
| $1^3F_4$ | 10.36 | 10.4164 | -0.0717 | -0.0001 | 0.0035 | 0.0250 | -0.0124 | 0.0126 |

## V. CONCLUSION AND OUTLOOK

We have constructed this model for heavy mesons based on a relativistic bound state equation, namely the spectator equation. The calculated spectra are in quite good agreement with the experimental data. The parameter values we have are reasonable, and comparable to other models of similar type. The model is derived by expanding the spectator equation in $1/M_Q$, where $M_Q$ is the mass of the heavy quark. This treatment is expected to work better for $b$-flavored mesons than for $c$-flavored mesons since in $c$-flavored mesons, $v \sim \frac{1}{2}c$, but in $b$-flavored mesons, $v \sim \frac{1}{5}c$, and our results confirm this expectation.

The retardation contribution to the $Q\bar{Q}$ mesons, which is missing in other quark models, has a noticeable effect. Annihilation effects have been neglected, as they are suppressed by additional powers of $\alpha_s(M_Q)$, which is a small parameter.

In addition to the questions currently being investigated (parametrization of $\alpha_s(r)$, $1/m_Q$ terms), this work opens up many avenues of investigation. Of primary importance is the application of the model to decay processes of heavy mesons. In particular, the calculation of the Isgur-Wise functions that describe the semileptonic decays, not only for decays to pseudoscalars and vectors, but also to excited states, are of great interest. In HQET, these form factors are essentially the overlaps of the appropriately boosted wave functions. It will be interesting to see if this relationship between the form factors and the wave functions arises in the present model, and if so, how. In addition, the slope of the Isgur-Wise function for the elastic decays may also be calculated, and various HQET sum rules checked.

The strong and electromagnetic decays may also be treated with the wave functions that we have. These are particularly interesting for the $D^*$ and $D_s^*$ states, as the former lie so close to the $D\pi$ threshold, while the latter lie below the $DK$ threshold, and thus decay radiatively. In addition, quantities such as meson decay constants may also be evaluated.

## ACKNOWLEDGEMENTS

The authors would like to acknowledge many useful conversations with Franz Gross and Nathan Isgur. This work was supported by the Department of Energy under contracts DE-AC05-84ER40150 and DE-FG05-94ER40832, and by the National Science Foundation under the National Young Investigator program.

[1] N. Isgur and M. B. Wise, Phys. Lett. **232B**, 113 (1989); **237B**, 527 (1990).
[2] N. Isgur and M. B. Wise, in *B Decays*, World Scientific (1992), Sheldon Stone, ed., p. 158.
[3] M. Neubert, Phys. Rep. **245**, 259 (1994).
[4] S. Godfrey and N. Isgur, Phys. Rev. D **32**, 189 (1985).
[5] F. Gross, Phys. Rev. **186**, 1448 (1969).
[6] F. Gross, J. W. Van Orden and K. Holinde, Phys. Rev. C **41**, R1909 (1990); C **45**, 2094 (1992).
[7] F. Gross and J. Milana, Phys. Rev. D **43**, 2401 (1991); D **45**, 969 (1992); CEBAF preprint CEBAF-TH-94-01.
[8] F. Gross, Phys. Rev. C **26**, 2203 (1982).
[9] W. W. Buck, Phd. Thesis, College of William and Mary, (1976), unpublished.
[10] D. Gromes, Nucl. Phys. **B131**, 80 (1977).
[11] M. G. Olsson and K. J. Miller, Phys. Rev. D **28**, 674 (1983).
[12] S. N. Mukherjee *et al.*, Phys. Rep. **231**, 201 (1993).
[13] De Rújula, Howard Georgi, and S. L. Glashow, Phys. Rev. D **12**, 147 (1975).
[14] J. D. Bjorken and S. D. Drell, *Relativistic Quantum Mechanics*, ( McGraw-Hill, New York, 1964).
[15] W. H. Press, B. P. Flannery, S. A. Teukolsky and W. T. Vetterling, *Numerical Recipes, The Art of Scientific Computing,* (Cambridge, New York, 1986).
[16] Review of Particle Properties,Phys. Rev. D **50** 3-I (1994)
[17] H. J. Schnitzer, Phys. Lett. **76B**, 461 (1978).